\documentclass[man,11pt,letterpaper,hidelinks,floatsintext]{apa7} 

\usepackage[american]{babel}
\usepackage{csquotes} 
\usepackage{url} 
\usepackage{doi} 

\usepackage{ragged2e} 
\usepackage{amsmath,amsthm,amssymb,dsfont}
\usepackage{booktabs}
\usepackage{tabularx}
\usepackage{tabulary}
\usepackage{multirow}
\newcolumntype{s}{>{\setbox0=\hbox\bgroup}c<{\egroup}@{}} 
\usepackage{caption}
\usepackage{subcaption}
\usepackage{etoolbox} 
\usepackage{setspace} 
\AtBeginEnvironment{table}{\singlespacing}
\AtEndEnvironment{table}{\doublespacing} 

\usepackage[style=apa, backend=biber]{biblatex}
\usepackage[separator = {;}]{credits}
\credit{PD} {1.0,0.0,0.0,1.0,0.0,0.5,0.0,0.0,0.0,1.0,1.0,0.0,0.75,1.0}
\credit{MF} {1.0,0.0,0.0,0.0,0.0,1.0,0.0,0.0,1.0,0.0,0.0,0.0,0.0,1.0}
\credit{VS} {0.5,1.0,1.0,0.0,1.0,1.0,0.0,0.0,0.5,0.0,1.0,1.0,0.75,1.0}
\credit{JS} {1.0,0.2,0.2,0.0,0.2,0.0,1.0,1.0,0.0,0.2,0.0,0.5,0.75,1.0}

\title{Interpretable Prediction Rule Ensembles in the Presence of Missing Data}
\shorttitle{Prediction Rule Ensembles in the Presence of Missing Data}
\authorsnames[1,1,2,1]{Vincent Schroeder*, Jakob Schwerter*, Marjolein Fokkema, Philipp Doebler}

\authorsaffiliations{{TU Dortmund University}, {University of Leiden}}

\authornote{ \small
   \noindent * Contributed equally and share the first authorship.

   \noindent Communicating author contact details: \addORCIDlink{Jakob Schwerter}{0000-0001-5818-2431}, TU Dortmund University, Department of Statistics, Martin-Schmeisser-Weg 13, D-44227 Dortmund, Germany. Email: \texttt{jakob.schwerter@tu-dortmund.de}. 
   
\noindent The authors gratefully acknowledge the computing time provided on the Linux HPC cluster at Technical University Dortmund (LiDO3), partially funded as part of the Large-Scale Equipment Initiative by the Deutsche Forschungsgemeinschaft (DFG, German Research Foundation) as project 271512359.

\noindent The project ``From Prediction to Agile Interventions in the Social Sciences (FAIR)'' has received funding from the programme ``Profilbildung 2020'' (PROFILNRW-2020-068), an initiative of the Ministry of Culture and Science of the State of North Rhine Westphalia. Sole responsibility for the content of this publication lies with the authors.

\noindent This study was not preregistered.

\noindent Author contributions: \insertcreditsstatement

\noindent Author contributions (weighted): \\ \insertcredits

}

\abstract{
Prediction Rule Ensembles (PREs) are robust and interpretable statistical learning techniques with potential for predictive analytics, yet their efficacy in the presence of missing data is untested. This study uses multiple imputation to fill in missing values, but uses a data stacking approach instead of a traditional model pooling approach to combine the results. 

We perform a simulation study to compare imputation methods under realistic conditions, focusing on sample sizes of $N=200$ and $N=400$ across 1,000 replications. Evaluated techniques include multiple imputation by chained equations with predictive mean matching (MICE PMM), MICE with Random Forest (MICE RF), Random Forest imputation with the ranger algorithm (missRanger), and imputation using extreme gradient boosting (MIXGBoost), with results compared to listwise deletion. Because stacking multiple imputed datasets can overly complicate models, we additionally explore different coarsening levels to simplify and enhance the interpretability and performance of PRE models. 

Our findings highlight a trade-off between predictive performance and model complexity in selecting imputation methods. While MIXGBoost and MICE PMM yield high rule recovery rates, they also increase false positives in rule selection. In contrast, MICE RF and missRanger promote rule sparsity. MIXGBoost achieved the greatest MSE reduction, followed by MICE PMM, MICE RF, and missRanger. Avoiding too-course rounding of variables helps to reduce model size with marginal loss in performance. Listwise deletion has an adverse impact on model validity. Our results emphasize the importance of choosing suitable imputation techniques based on research goals and of advancing methods for handling missing data in statistical learning.
}

\begin{document}

\maketitle


\section{Introduction} 

Empirical studies in fields such as psychology and education have traditionally focused on a limited set of theorized predictors, largely due to methodological constraints. While traditional regression methods facilitate confirmatory hypothesis testing, they limit the number of predictor variables that can be included \parencite{ByrnesM2007}. Machine learning methods, especially those with built-in feature selection, allow for the inclusion of a large set of potential predictor variables while still providing interpretable results. This is important, for example, in research areas such as learning analytics, where various log (and survey) data are used to predict outcomes such as student success. In these cases, the first goal is to achieve high prediction accuracy, and next to understand which features were most helpful for prediction \parencite{ArizmendiEA2022}.

Popular methods for interpretable machine learning include penalized regression, such as Lasso or Elastic Net and/or tree-based methods such as Random Forest and XGBoost \parencite{ParkEA2023, LupyanG2019, ArizmendiEA2022}. Although Random Forest and XGBoost show promise in improving prediction accuracy over Lasso, beyond variable importances, additional effort is required for detailed interpretation, especially to understand the functional shape of relations and variable interactions. Prediction rule ensembles \parencite[PRE; ][]{FriedmanP2008, FokkemaS2020} combine the strengths of (penalized) linear regression with ensembles of decision trees. Ensembles of decision trees, such as Random Forest, are powerful prediction methods, but PREs additionally aim to improve interpretability. The method not only allows for incorporating a large number of potential predictor variables, but also for capturing their linear, non-linear as well as interaction effects. As such, it allows for understanding not only the effects of individual predictors, but also how variables interact with each other, potentially in non-linear ways. PREs are therefore a promising interpretable machine learning method for social science research.

Research and applications require strategies  for handling missing data, which is often unavoidable when collecting behavioral data. Random missingness can often be compensated for by multiple imputation \parencite{BuurenG2011} followed by averaging over all imputed data sets using Rubin's rule \parencite{Rubin1986}. In machine learning, especially with feature selection (such as in PRE), this procedure is not applicable: When applying variable selection, the set of selected variables depends on the specific imputed dataset, and the set will vary across multiple imputations. In response to this problem, stacking all imputed datasets into one large dataset has been proposed and shown to be effective for penalized regression \parencite{DuB2022, ThaoG2019}. However, it is unclear how different imputation methods may affect the performance of PREs. Furthermore, it has been pointed out that the handling of missing data in machine learning prediction studies is often inadequate, with researchers resorting to suboptimal approaches such as complete cases analyses \parencite{NijmanL2022,navarro2023systematic}. Therefore, we present a simulation study on analyzing multiple imputed data using PRE. The results will guide users in dealing with missing data in various data-analytic situations.

Our study design incorporates different imputation methods in a fully crossed simulation design, with different amounts of missingness and sample sizes. 
Furthermore, our simulation study investigates whether coarsening may improve performance when analyzing multiple imputed data with PRE. PRE results may become overly complex because of the method's tendency to repeatedly select the same variable or combination of variables with different cut-off values. This is exacerbated whenever the number of unique predictor variable values increases, as is generally the case with multiple imputation. We evaluated the effect of different levels of coarsening in our simulation by rounding variables, e.g., to the first decimal place or quantiles. Coarsening aims to enhance model simplicity and interpretability by ensuring that the model does not become overly complex. 
\subsection{A brief introduction to prediction rule ensembles}
Prediction Rule Ensembles (PREs) were originally developed by \textcite{FriedmanP2008}. Several extensions to the methodology have been developed by \textcite{Fokkema2020}, such as generalizations to response variable types other than continuous and binary outcomes. A prediction rule ensemble combines linear effects and rules in a single model. These rules can be expressed as relatively simple conditional statements, e.g.,  ``if conditions A and B are met, then modify the prediction by $\alpha$''. Each rule is similar to a decision-making process and can also be visualized as a very simple decision tree. In contrast to classical confirmatory methods, neither conditions (A,B)  nor coefficients ($\alpha$) need to be known a-priori, but PRE constructs a large set of potential rules using a tree ensemble, from which a subset are selected and corresponding coefficients estimated using a penalized regression method, such as the Lasso. PREs are versatile and capable of handling varied data types including numeric, categorical, and count variables. PREs typically select multiple rules and linear effects, forming the final rule ensemble. In practice, relatively few rules are often sufficient for good predictive performance, sometimes even fewer than ten \parencite{FokkemaS2020}.

Following \textcite{FriedmanP2008} and \textcite{FokkemaS2020}, the ensemble's predictive model is given by
\[
y_{i} = \alpha_0 + \sum_{k=1}^{K} \alpha_k r_k(x_i) + \sum_{j=1}^{J} \beta_j l_j(x_{ij})  + \epsilon_i\text{,}
\]
where $y_i$ is the $i$th individual's outcome, $r_k(x_i)$ denotes the $k$th rule derived from covariates $x_i$ (with $r_k(x_i) = 1$ if the rule applies and $r_k(x_i) = 0$ otherwise), and $l_j(x_{ij})$ is a winsorized version of the $j$th covariate. Winsorizing is used to reduce the effect of outliers, and sets values beyond the $q$th and $(100-q)$th quantile of a variable to the value of that quantile \parencite{MairW2020}\footnote{By default, $q$ is set to 0.025.}. PRE thus captures linear main effects of the predictors through the linear terms $\beta_j l_j(x_{i})$, while it captures non-linear and interaction effects through the rules $\alpha_k r_k(x_i)$. 

Although it is possible to incorporate a-priori known rules in the ensemble, the rules are generally unknown prior to the analyses, and therefore need to be derived from the dataset. In principle, any tree ensembling method can be used, such as Random Forest or (extreme) gradient boosting. By default, an ensemble is created using gradient boosting, and the implementation by \textcite{Fokkema2020} employs the conditional inference tree algorithm \parencite{HothornH2006} for fitting the trees, which selects splitting variables in an unbiased manner. Specification of the maximum tree depth controls the complexity of rules by limiting the order of interactions in each rule and hence the final rule ensemble.

Coefficients $\alpha_0, \dots, \alpha_K$ and $\beta_1, \dots, \beta_J$ are found by minimizing the loss function, given by
\[
\sum_{i=1}^{N} L\left(y_i, \alpha_0 + \sum_{k=1}^{K} \alpha_k r_k(x_i) + \sum_{j=1}^{J} \beta_j l_j(x_{ij})\right),
\]
a measure of discrepancy between the predictions and observations. In the case of a continuous dependent variable, $L$ will often be the squared error loss, but PREs can handle various dependent variable types available within the generalized linear model (GLM) with other loss functions. 

To penalize model complexity and curb overfitting, a penalty term is added to the loss function. For example, the Lasso penalty term
\[
\lambda \left(\sum_{k=1}^{K} |\alpha_k| + \sum_{j=1}^{J} |\beta_j|\right),
\]
where the tuning parameter $\lambda$ governs the degree of sparsity, i.e., the proportion of non-zero coefficients. Other sparse or penalized regression methods may also be employed. We use a variant of the Lasso penalty here: After the regular Lasso penalty has performed model selection, the relaxed Lasso technique \parencite{HastieT2020} re-estimates non-zero coefficients to avoid overly conservative shrinkage of large coefficients.

Both coefficients of rules and linear terms can be interpreted as well-known GLM coefficients. To aid in interpretation, importances are produced, which are the product of the absolute values of the coefficients and the sample standard deviations of the linear term or rule. Thereby, relative contributions can be compared between rules and linear terms. The importance of the $j$th linear term is given by
\[
  I_j = |\hat{b}_j| \cdot \mathit{sd}(l_j(x_j)),
\]
while the importance of the $k$th rule is given by
\[
  I_k = |\hat{a}_k| \cdot \sqrt{s_k(1-s_k)},
\]
where $s_k$ is the support of rule $k$ in the training data, that is, the proportion of training observations to which rule $k$ applies.

Finally, the importance of input variable $x_j$ is given by the sum of the importances of the linear term and the rules in which $x_j$ appears:
\[
  J_j = I_j + \sum_{x_j \in r_k}\frac{I_k}{c_k},
\]
where $c_k$ is the number of input variables in rule $k$. Note that the importance of a rule is thus distributed equally over the input variables defining the rule. If a variable appears more than once in the conditions of a rule, $I_k / c_k$ would be multiplied accordingly. All importances take values $\geq 0$, with higher values indicating a stronger effect on the ensemble's predictions.

To aid in the interpretation of the rule ensemble, partial dependency plots provide a visual representation of the effect of a single variable or pair of variables on the outcome, holding all other variables constant. This methodological approach is facilitated by the convenient R package \texttt{pre} \parencite{Fokkema2020}. 

\section{Current challenges: Missingness and Coarseness}
\subsection{How should one coarsen data?}
A lot of rules may potentially be created, and the number of possible rules increases with the number of unique values that are observed for each variable. While predictive performance potentially profits from many different splitting values for the same variable because predictions can be more fine-grained, in terms of interpretation, a smaller number of unique rules may be preferable. Prior research has not yet studied this trade-off between predictive performance and coarsening variables (i.e., by rounding to the first decimal place or rounding to quantiles), so a systematic investigation is in order. 

\subsection{How should one impute missing data?}
Some 20 years ago, \textcite{SchaferG2002} recommended full information maximum likelihood or (Bayesian) multiple imputation as general approaches to deal with missing data. As pointed out by \textcite{Enders2023}, this advice may not similarly apply in predictive modeling contexts, where the goal is often to discover non-linear or interactive effects, instead of fitting confirmatory models that allow for hypothesis testing. \textcite{Enders2023, NijmanL2022} explicitly call for methodological innovations in this area. This paper responds to this call by evaluating the performance of different imputation methods for managing missing data in machine learning prediction models under different sample sizes and missing-data scenarios.

A typical MI workflow generates five or more imputed datasets. Five separate analyses are then run and subsequently pooled using Rubin's rules \parencite{WoodW2008}. In the case of PREs, the typical MI workflow potentially results in five different sets of selected rules due to the Lasso selection, which do not necessarily overlap, limiting the interpretability of the results. Therefore, a different MI strategy is required in the case of feature selection \parencite{WoodW2008}. We adopt the stacking approach of \textcite{WoodW2008, DuB2022, ThaoG2019}, which combines all imputed datasets into one. That is, five datasets are combined into one dataset, with five times the sample size. The optimal value of the Lasso penalty parameter is in principle not sensitive to sample size, so the artificially increased sample size will likely not increase overfitting. Selection of the optimal value of the Lasso penalty, however, is performed by $k$-fold cross-validation. This may result in overly optimistic penalty parameter values (i.e., too little penalization) if the observations in the different folds are dependent. A key characteristic of our approach is therefore to avoid this over-optimism by assigning the multiple imputations of the same observation to the same fold when choosing the optimal value of the penalty parameter.

For Lasso regression, \textcite{DuB2022} compare the stacking approach with a so-called ``grouped'' approach in which separate models are fitted to each imputed dataset, the objective functions of the models are pooled, and the optimization is performed jointly over the pooled objective function. Their simulations show that the stacked approach is more computationally efficient and has better estimation and selection properties, a conclusion shared by \textcite{ThaoG2019}. \textcite{GunnH2023} analyzed a single dataset and found slightly better predictive accuracy on test data for the pooled compared to the stacked approach, but the difference in proportion of variance explained was $<0.01$. It is unclear whether \textcite{GunnH2023} took into account the dependency of imputations of the same observations when selecting the penalty parameter. They explain the CV used to optimize lambda in great detail, but do not mention the problems of assigning multiple imputations of the same observation to the same fold. This may explain why their stacked lasso model selects so many variables while the other methods do not. Furthermore, this finding is in line with \textcite{WoodW2008}, who found that the stacking approach performs much better than listwise deletion while providing a good approximation of the gold standard of variable selection by repeated use of Rubin's rule, at much lower computational cost.

\section{Simulation Study}

\subsection{Method}

\subsubsection{Data generation}
Each condition of the simulation is repeated 1,000 times. In each repetition, pseudo-random samples from 10 standard-normally distributed variables are drawn. The first four variables $X_1$ through $X_4$ have a pairwise correlation of 0.3, variables five and six are negatively correlated with -0.25 and the other variables are uncorrelated. This inter-correlation between the predictors is necessary to ensure that imputation of missing values is possible.

The response variable $Y$ is a function of one linear term, six rules and an error term, which corresponds to realistic model output in real-world data scenarios \parencite{FokkemaS2020}. These terms and their coefficients are presented in Table~\ref{tab:simulation_terms}. The first term (\#1) represents a linear effect: For a unit increase in the (winsorized) value of $X_7$, the response variable increases by $\beta_1 = 0.5$. Note that $l({X_7})$ refers to the $95\%$ winsorized predictor $X_7$. The six rules represent non-linear interaction effects of predictors $X_1$, $X_2$, $X_5$, $X_6$, $X_8$, $X_9$ and $X_{10}$. For example, rule \#3: $X_{10} > 0.5\ \& \ X_1 > 0.5$ implies that for all observations which have  values $X_{10}$ and $X_1$ larger than the cut-off value 0.5, the response variable $Y$ increases by the true coefficient of 1. 

Higher cut-off values result in fewer observations that meet the conditions of the rule, as can also be seen in the prevalence column of Table~\ref{tab:simulation_terms}.
Larger coefficients lead to larger increases in the response variable and therefore indicate a greater importance of the rule.
Finally, the noise term $\epsilon$ follows a standard normal distribution and the full data generation process is given by

\begin{equation}
y_{i} = \beta_1 l(x_{i7}) + \sum_{k=2}^{7} \alpha_k r_k(x_i) +  \epsilon_i.
\end{equation}

\begin{table}[ht!]
\begin{center}
\caption{All rules and linear terms in the simulation model.}
    \begin{tabular}{rlccc}
  \hline
$\#$  & variable(s) and cut-off values & type & true coefficient & prevalence \\ 
  \hline
1 & $X_7$ & linear  & 0.50 & - \\ 
  2 & $X_9$ $>$ -1 \& $X_2$ $>$ -1 & rule & 0.75  & 0.702\\ 
  3 & $X_{10}$ $>$ 0.5 \& $X_1$ $>$ 0.5 & rule & 1.00 & 0.094 \\ 
  4 & $X_{10}$ $>$ -0.75 \& $X_5$ $>$ -0.75 & rule & 1.25 & 0.600  \\ 
  5 & $X_8$ $>$ 0.25 \& $X_1$ $>$ 0.25 & rule & 1.50 & 0.161 \\ 
  6 & $X_6$ $>$ -0.5 \& $X_5$ $>$ -0.5 & rule & 1.75 & 0.445 \\ 
  7 & $X_1 $$>$ 0 \& $X_2$ $>$ 0 & rule & 2.00 & 0.299\\ 

   \hline
\end{tabular}
\label{tab:simulation_terms}
\end{center}
\textit{Notes:} prevalence: probability that the $k$th rule applies to a random synthetic observation.
\end{table}

We varied three factors, each with two levels, resulting in $2^3 = 8$ data generating conditions: sample size ($N = 200$ or $N = 400$), rate of missingness (0.08 or 0.48), and missing-data mechanism (MCAR or MAR). Note that imputation and coarsening further changes the data, but the same datasets were imputed and coarsened in different ways. The complete data was stored and analyzed as a benchmark, and also analyzed to evaluate different kinds of rounding.

Under MCAR, each row is equally likely to contain missing values. The response variable is never missing. With an overall missing rate of 8\%, 40\% of the rows have two variables missing. All combinations of variables are possible and equally likely to be missing. When the missing rate is 48\%, each predictor has 48\% missing values. That is, 80\% of the rows were randomly selected to contain missings for a random selection of 60\% of the predictor variables. The MCAR amputation is done via function \verb|ampute| from package \verb|mice| \parencite{BuurenG2011}.

The MAR amputation algorithm is based on \textcite{thurow_goodness_2021} and has been slightly adapted to fit the simulation here.
The probability of a value $x_{ij}$ to be missing, depends on the value of the tenth predictor variable $x_{i10}$. The values $x_{1,10}, ..., x_{N,10}$ are ordered and then discretized into 20 intervals of equal length. Within each interval, the missing probability is constant. That way, low values of $X_{10}$ lead to high probabilities for all other predictor variables to be missing for a given observation, with a stepwise increase in the missingness probability. $X_{10}$ itself is never missing to make sure the mechanism is not MNAR. For comparability to the MCAR condition, $X_{10}$ is also never missing. The overall missingness proportion is the same under MAR and MCAR, but under MAR the number of variables missing for a row can vary, while under MCAR, it is fixed.

\subsubsection{Imputation and coarsening}
We varied three factors:
\begin{itemize}
\item Missing data handling (5 levels): Listwise deletion (LD; no imputation) as a baseline, multiple imputation by chained equations (MICE) using predictive mean matching \parencite[PMM; ][]{Rubin1986},  MICE using Random Forest \parencite[RF; ][]{breiman2001random}, Random Forest using the ranger algorithm  \parencite[missRanger; ][]{missRangerP} or imputation using extram gradient boosting  \parencite[MIXGBoost; ][]{mixgbP}.
\item Number of imputations (2 levels): one or five.
\item Coarsening (4 levels): No rounding, rounded to one decimal, decile quantization or quintile quantization.
\end{itemize}

MICE \parencite[]{miceP} is a robust method which imputes missing data in a iterative way, one variable at a time. Each variable containing missing values acts as the dependent variable to get predictions for it, using the other variables in the dataset as predictors. The predictions generated by these models are then used to fill in missing values and will be used as predictors for the next variable to be imputed, for which a new prediction model is trained. This process is repeated multiple times, until convergence is met. A large range of methods for the prediction models are supported by the R package \verb|mice|. In this simulation, Predictive Mean Matching (PMM) with 5 prediction donors and Random Forests were used. 

PMM is an effective semi-parametric imputation method that involves calculating predicted values for each target variable with missing entries. Instead of imputing directly from the conditional distribution of the missing variable, PMM forms a donor pool from fully observed data, where donors are selected based on the similarity of their predictions to the missing variable. An imputation value is randomly selected from these donors, which helps to preserve the distributional properties of the original data and ensures that the imputed values are plausible \parencite{vanBuuren2018}. It has been found that using smaller donor pools, such as the default of five in \verb!mice! \parencite{vanBuuren2018}, leads to more accurate results than larger pools \parencite{Kleinke2017}. Essentially, instead of using arbitrarily selected or averaged observations, PMMs use observations from the same variable or distribution which most closely match the predictions. 

The Random Forest algorithm, introduced by \textcite{breiman2001random}, is a versatile machine learning method that uses an ensemble of decision trees to effectively handle complex relationships and high-dimensional data. Each tree in the Random Forest acts as either a classifier or a regressor, depending on the task, and contributes to the final decision either by voting (in classification) or by averaging predictions (in regression). This ensemble approach increases the accuracy and robustness of the model while reducing the risk of overfitting. Random Forests consider only a random subset of features at each split in a tree, increasing the randomness and diversity of the individual trees, which beneficially affects performance of the full ensemble \parencite{breiman2001random, CutlerEA2012}. MICE RF begins by partitioning the dataset based on observed and missing values, and uses Random Forest to iteratively predict missing entries.

MissRanger \parencite[]{missRangerP} combines the Random Forest using the ranger algorithm \parencite{RangerWright2017} and the PMM aproach. Random Forest models are trained sequentially to generate predictions for each variable with missing data. However, instead of directly using these predicted values for imputation, missRanger uses these predictions in the next step involving PMM to get predicted values which are already present in the data and to achieve appropriate imputation variability. 

Finally, MIXGBoost \parencite[]{mixgbP} uses the Extreme Gradient Boosting (XGBoost) algorithm, which combines gradient boosting with regularization techniques to prevent overfitting and improve model performance \parencite{ChenG2016} to impute missing data. MIXGBoost extends these capabilities specifically for imputation by incorporating features such as subsampling and predictive mean matching (PMM) that are tailored to effectively handle different types of data. In the MIXGBoost workflow, the dataset is first split based on the presence of observed and missing values. The algorithm performs an initial imputation to create a complete dataset, which is then used to train MIXGBoost models under different subsampling ratios. This approach allows the models to effectively capture the uncertainty associated with missing values. For continuous variables, PMM is used with a standard number of donors, which improves the accuracy of the imputations \parencite{DengL2023, SuhS2023}.

In the case of multiple Imputations (MI), the imputation process is independently repeated 5 times. The five resulting datasets are stacked into one long dataset.

After LD or imputation, the data was coarsened to one of four possible degrees: No rounding, rounding to one decimal, decile quantization and quintile quantization. Since the predictor variables were standard-normally distributed, rounding to one decimal can be interpreted as coarsening by 0.1 standard deviations and leads to fewer unique values. For example, a sample consisting of 400 unique standard normally distributed observations is reduced to approximately 51 unique values by rounding the variables to one decimal. Quantization to deciles results in ten unique values for each variable, and quantization to quintiles results in five unique values. Quantized values were set to the midpoint of the quantiles.

\subsubsection{Model fitting}
PREs were fitted using the package \verb|pre| \parencite{Fokkema2020}. For the most part, the default settings were used: 500 initial Conditional Inference Trees \parencite{HothornH2006} were trained on a random subset containing 50\% of the sample, boosting was applied with a learning rate of 0.01 and the value of the $\lambda_{1se}$ tuning parameter was found using 10-fold cross-validation. A few settings differed from the default: All coefficients were constrained to be positive and complementary rules were retained to ensure this constraint did not weaken predictive performance \parencite{FokkemaS2020}. That way, the final model finds the rules with the same direction of inequalities instead of selecting a reverse rule with a negative coefficient, keeping the interpretation simple. Maximum tree depth was set to two, so that no rules corresponding to three-way and higher-order interactions would be derived. The Lasso was relaxed, i.e., all non-zero coefficients after Lasso variable selection were re-estimated without penalization. This procedure avoids the Lasso's shrinkage and hence decreases bias in the non-zero coefficient estimates \parencite{HastieT2020}.

\subsubsection{Performance evaluation}
A multi-criterial evaluation of PRE performance after imputation and coarsening sheds light on the various trade-offs. Criteria include rule recovery (including true and false positives), coefficient bias, cut-off-value recovery, predictive performance and model size. With these criteria, the performance of the various imputation and coarsening strategies can be assessed relative to the complete data set on the one end and LD on the other. In general, the results for the complete dataset are expected to be the best in all performance evaluations. For the imputation methods, a result as close as possible to the complete dataset is preferable. Below, all criteria are described. 
 
\textit{Rule recovery and coefficient bias}. Since the data-generating model is a PRE, accurate recovery of the true linear and rule-like terms is a meaningful criterion. We count a rule as `recovered' if the model finds both variables that are included and the direction of the inequality is correct for both variables. For a linear term, it only needs to be selected by the model since there are no inequalities to be correct. A term that is selected according to these criteria but not among the true rules is classified as a false positive. A model can select multiple instances of the same term with different cut-off values and coefficients. For all instances of a recovered term, the coefficients are summed up to a total coefficient. The difference between the total coefficient and the true coefficient will be called coefficient bias. 

\textit{Cut-off-value recovery}. The cut-off distance describes how accurately the model finds the true cut-off values of the rule-like terms and is derived in the following manner: First, calculate the difference between the empirical rule cut-off value and true cut-off value for all recovered rules. Second, average over the two variables in the rule. Then, calculate the mean of these averages across all instances of the recovered rules and across all replications. Lastly, average over all rules which have been recovered at least once. The cut-off distance is conditional on recovery. 

For example, if the model contains the empirical rule $X_{10} > 0.4\ \& \ X_1 > 0.3$ with a coefficient of 0.3 and another rule $X_{10} > 0.2\ \& \ X_1 > 0.4$ with the coefficient 0.5, the true rule (\#3) $X_{10} > 0.5\ \& \ X_1 > 0.5$ counts as recovered, since there is at least one instance of a rule in the model which contains the two variables $X_{10}$ and $X_1$ and both inequalities are correctly identified. The coefficient bias over the two instances amounts to $1.0 - (0.5 + 0.3) = 0.2$ and the  cut-off distance to $(\frac{|0.4-0.5| + |0.3-0.5|}{2} + \frac{|0.2-0.5| + |0.4-0.5|}{2})\ /\ 2 = 0.175$.

\textit{Predictive performance}. The predictive performance is evaluated by the mean squared error (MSE) computed over 10,000 synthetic test observations, unknown to the model in the training stage. We take the mean squared difference between observed and predicted values:
        \[
        MSE = \frac{1}{10,000} \sum_{i=1}^{10,000}(y_i-\hat y_i)^2.
        \]
\textit{Model size}. Additionally, the size of the model is considered, which equals the number of linear terms and empirical rules selected by the model. Smaller models can be interpreted more easily.

\subsection{Results: Relative performance of imputation methods}
We first discuss results for MCAR, followed by the MAR results, which are very similar. 

\subsubsection{Rule Recovery}

\begin{figure}[H]
    \centering
    \caption{Rule recovery with different imputation methods, under different rates of missingness (MCAR) and under different data coarsening settings.}
    \includegraphics[width=1.2\linewidth]{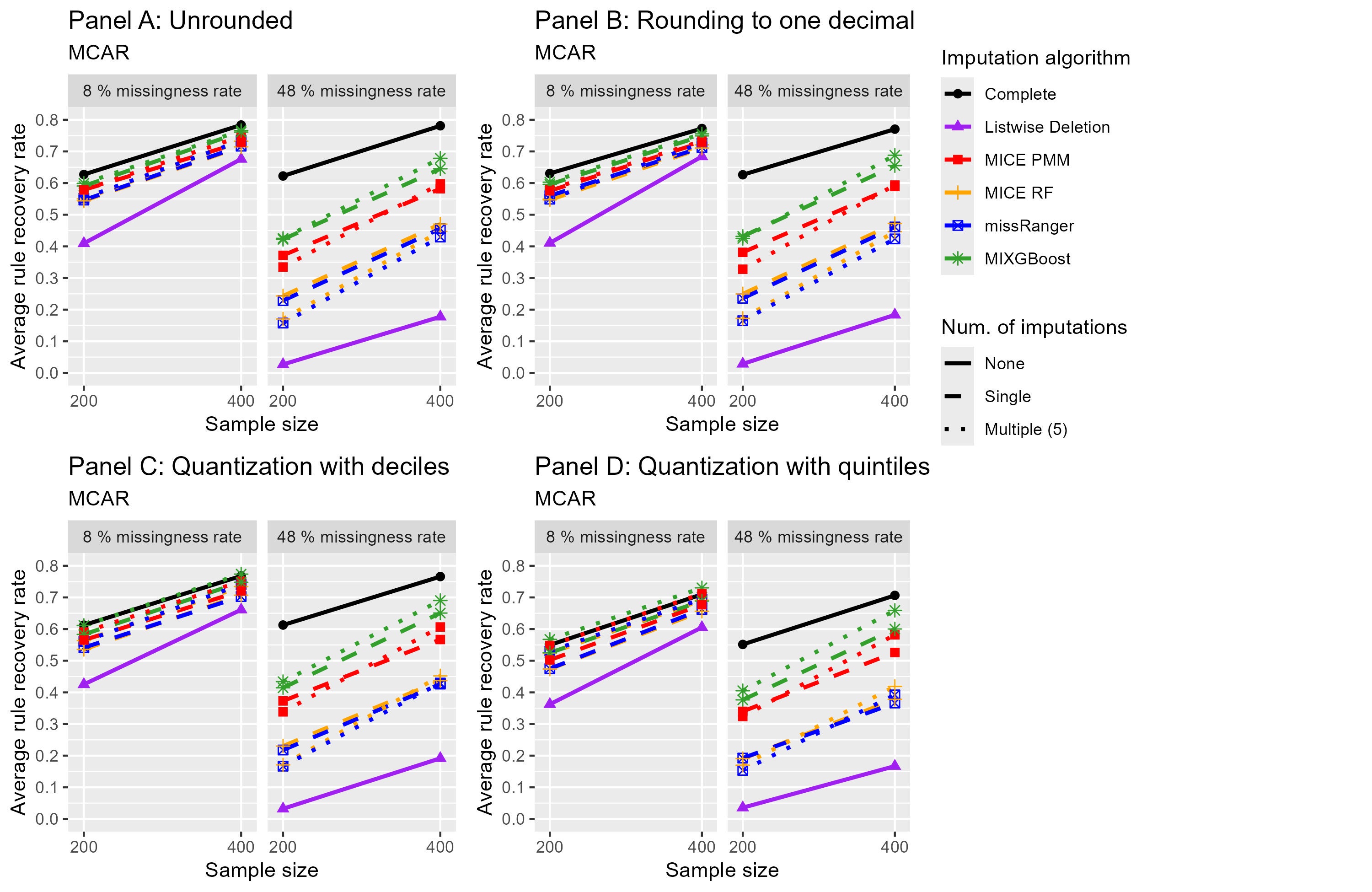}
    \label{fig:recovery}
\end{figure}

The ability to recover the true terms by the models under MCAR is shown in Panel A of Figure~\ref{fig:recovery}. In Panel A, the case without coarsening is considered. As expected, rule recovery is highest for complete data in almost all settings. This performance can be seen as a benchmark what is maximally achievable by the imputation algorithms. Also as expected, all imputation algorithms perform better than LD in all settings. In the most extreme case, when the sample size is small and missingness is high, LD fails to recover any of the rules and only recovers the linear term (see Table~\ref{tab:RR_LDvsImp} in the \hyperref[sec:Appendix]{Appendix}).

In nearly all settings, MI with MIXGBoost achieves the highest rule recovery compared to the other imputation algorithms. The largest gains in performance compared to LD are obtained when the sample size is large ($N=400$) and the missingness rate is high (48\%). As expected, with a lower missingness rate (8\%), MI performs better than single imputation for all imputation methods. However, with a higher missingess rate (48\%), single imputation performs slightly better or similar to MI for missRanger and MICE RF.

\subsubsection{Predictive Performance and Coefficient Bias}
\begin{figure}[H]
    \centering
    \includegraphics[width=1.2\linewidth]{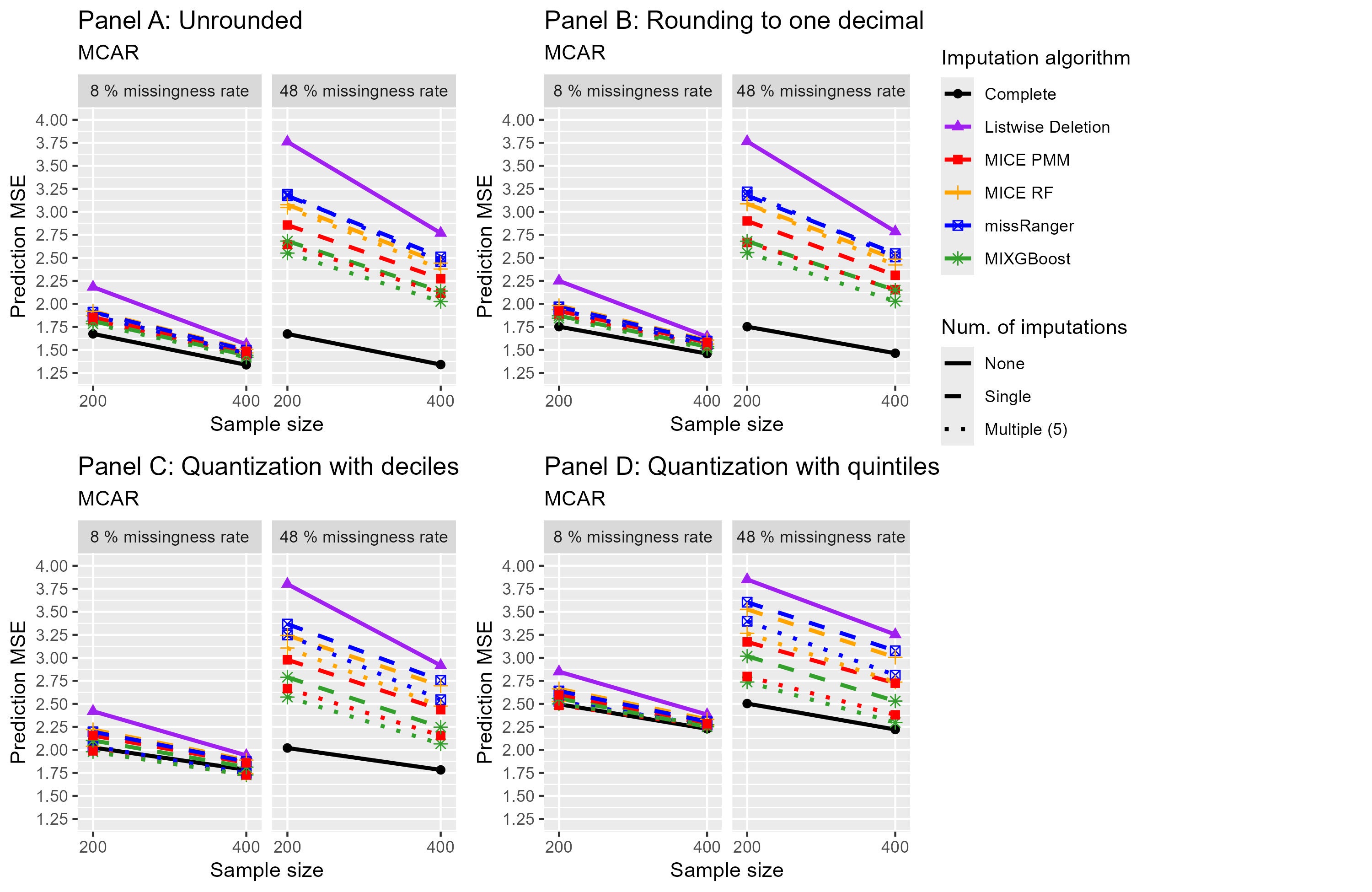}
    \caption{Predictive performance evaluated by the mean squared error (MSE) with different imputation methods, under different rates of missingness (MCAR) and different data coarsening settings.}
    \label{fig:MSE}
\end{figure}

Panel A of Figure~\ref{fig:MSE} shows that, as expected, all imputation methods achieve lower test MSE than LD. Between the imputation methods, MI with MIXGBoost is closest to the performance of the complete data. In most cases, MI outperforms SI, with the exception of missRanger, which shows similar MSE between MI and SI when the missingness rate is high.

\begin{figure}[H]
    \centering
    \includegraphics[width=1.2\linewidth]{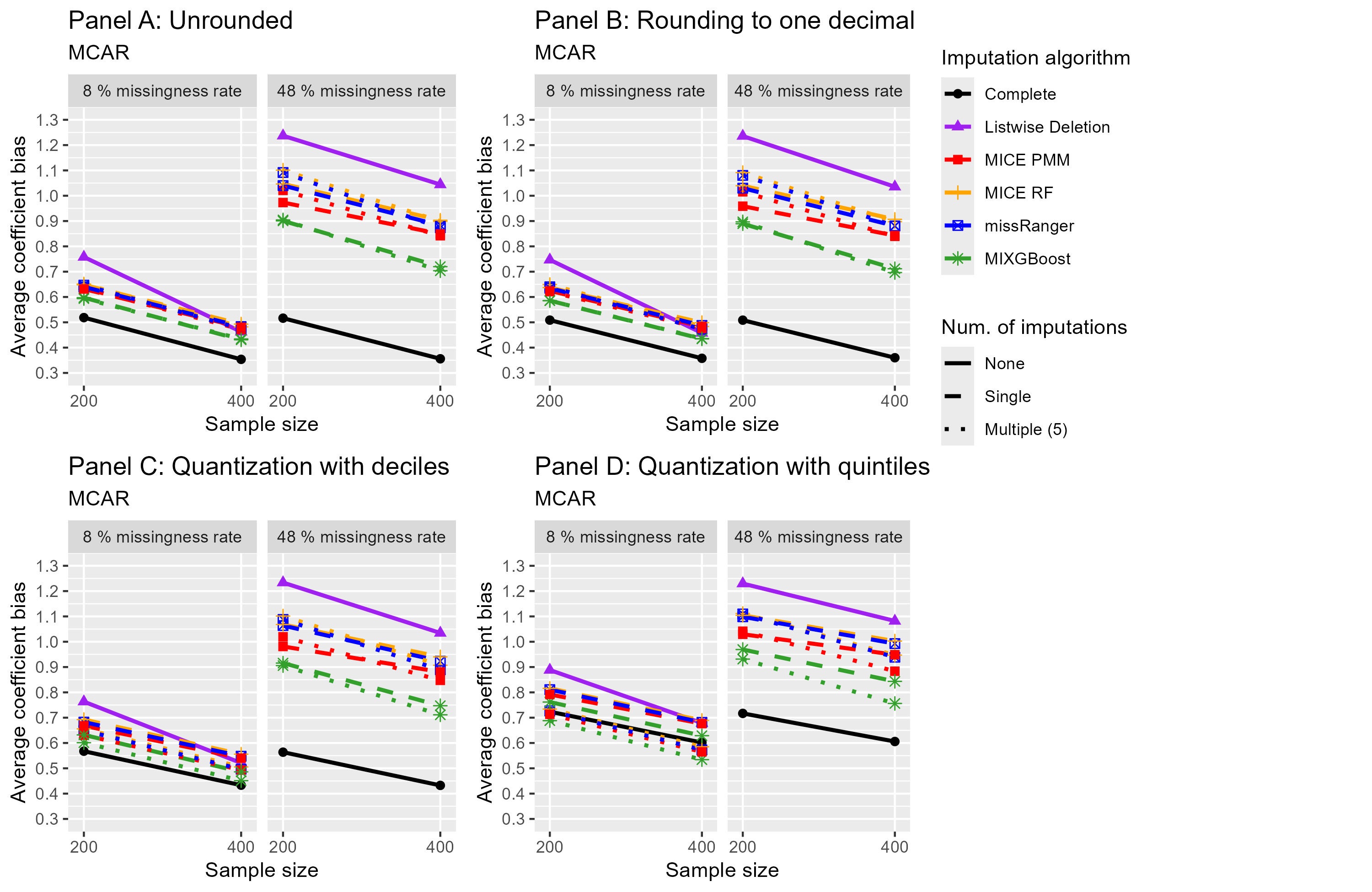}
    \caption{Average coefficient biases with different imputation methods, under different rates of missingness (MCAR) and different data coarsening settings.}
    \label{fig:bias}
\end{figure}

In part, good predictive accuracy is achieved by low bias in coefficient estimation. The coefficient bias achieved by LD and the imputation algorithms is presented in Panel A of Figure~\ref{fig:bias}. In all cases, MIXGBoost leads to the smallest average coefficient biases. When missingness is low and the sample size is large, LD is competitive and achieves about the same biases as MICE PMM, missRanger and MICE RF. In all other data settings, LD performs substantially worse than imputation. Between SI and MI, there is no clear advantage for either.

\subsubsection{False Positives and Model Size}

\begin{figure}
    \centering
    \includegraphics[width=1.2\linewidth]{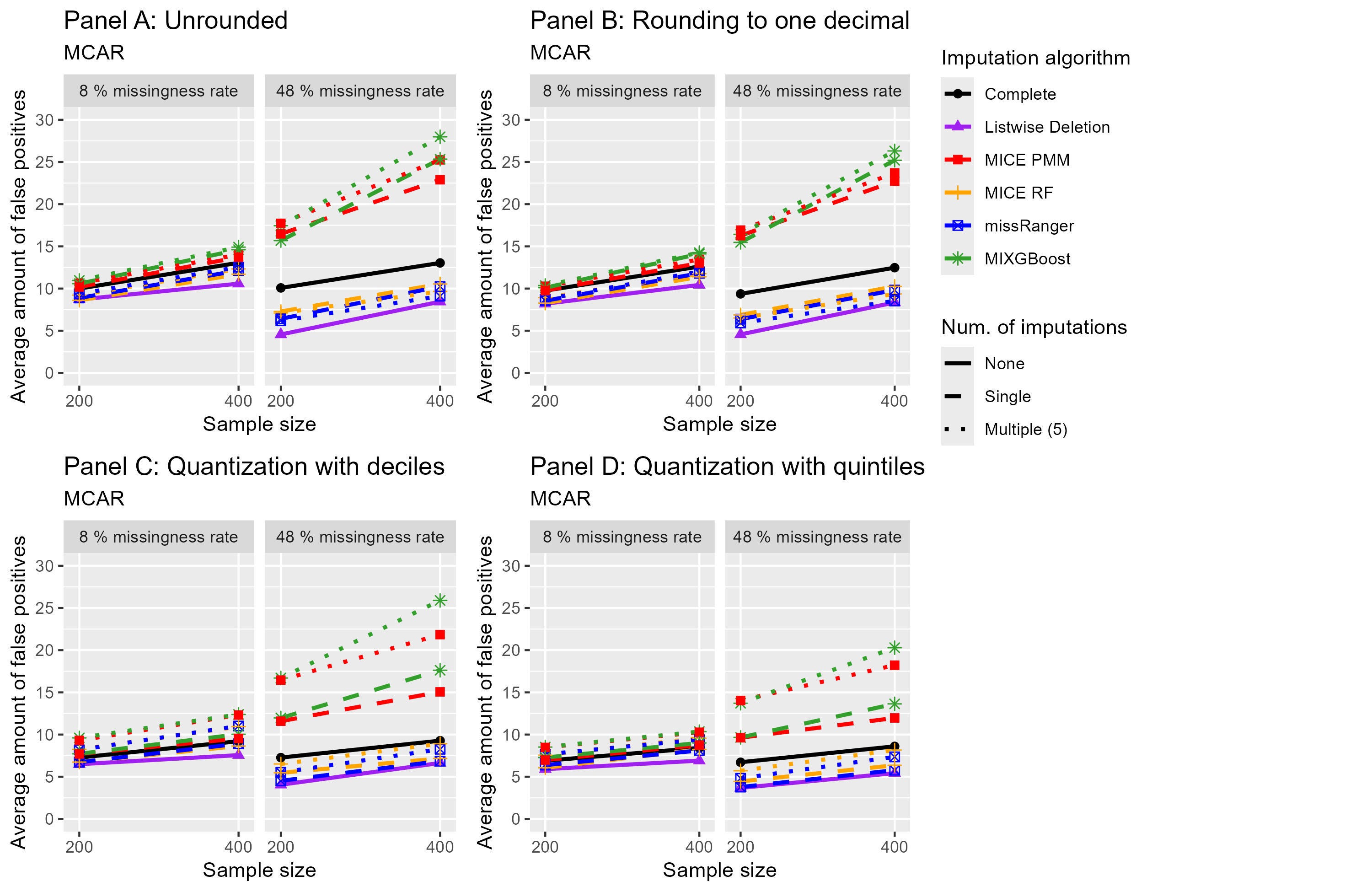 }
    \caption{Average number of false positives with different imputation methods, under different rates of missingness (MCAR) and different data coarsening settings.}
    \label{fig:FP}
\end{figure}

The number of selected false positives in the different settings is presented in Panel A of Figure~\ref{fig:FP}.
The models created using LD, missRanger and MICE RF select fewer false positives on average than models created on the complete data set. In contrast, using MIXGBoost and MICE PMM results in more false positives than on the full data set. This holds true for all levels of missingness and sample sizes. In most cases, using MI leads to more false positives than using SI, except for missRanger and MICE RF when missingness is high.

\begin{figure}[H]
    \centering
    \includegraphics[width=1.2\linewidth]{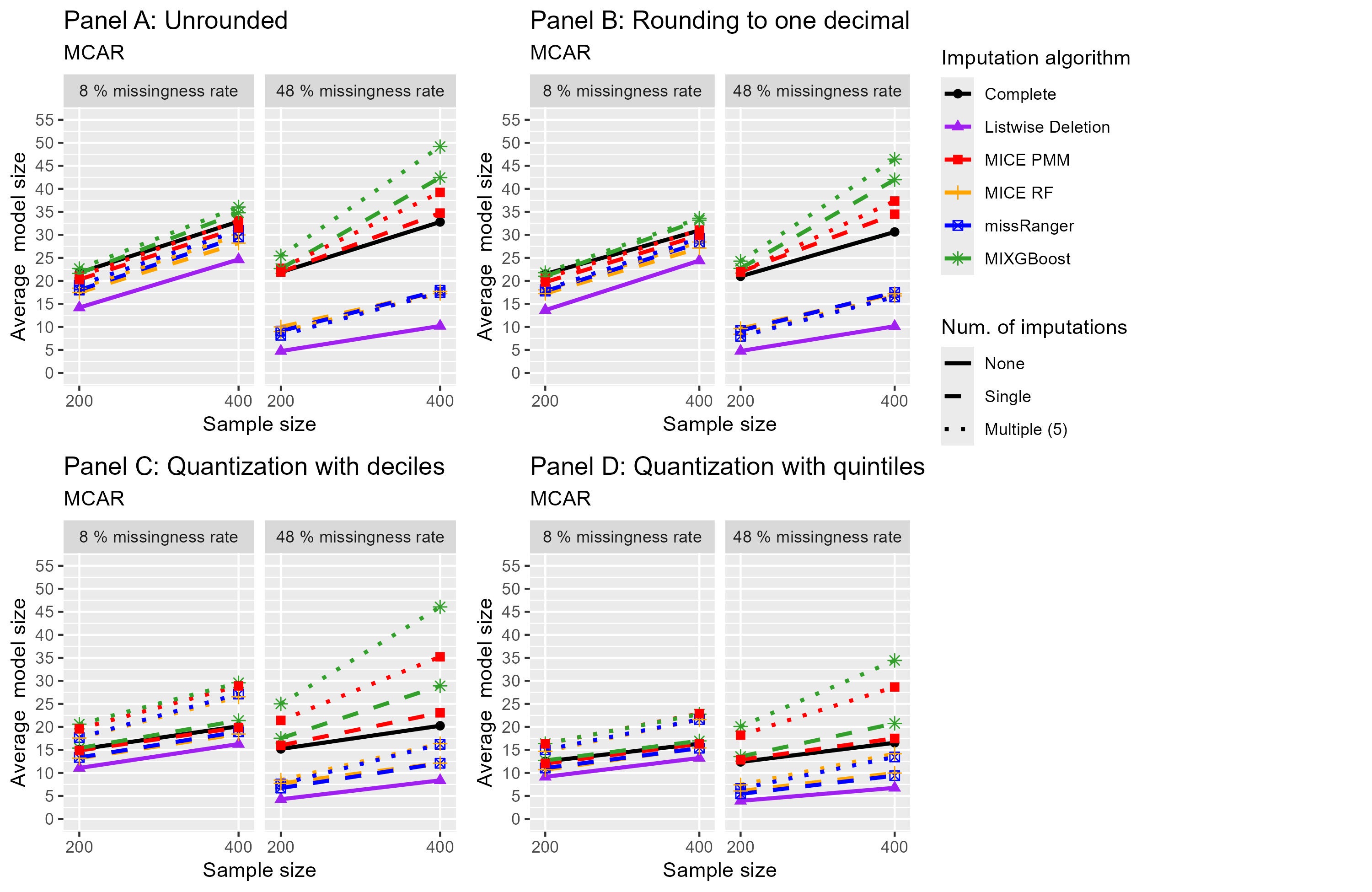}
    \caption{Average model size with different imputation methods, under different rates of missingness (MCAR) and different data coarsening settings.}
    \label{fig:Size}
\end{figure}

The number of false positives is positively correlated with total model size, i.e. the number of selected rules and linear terms, as shown in Panel A of Figure~\ref{fig:Size}. More false positives means that the model is larger and therefore less interpretable. LD leads to the smallest number of rules, possibly an effect of decreased sample size. For example, when missingness is high and the sample size is large, LD leads to models which contain only 10 terms on average, while models using the full data set contain 33 terms. This shows that LD results in overly conservative variable selection due to the loss of information. MIXGBoost MI yields the largest models in all cases, e.g., with high missingness and large sample size, 49 terms on average. The increased number of false positives when using MIXGBoost MI increases the total model size to 16 more terms on average, compared to the full data set. Using missRanger and MICE RF lead to smaller models than with the complete data set and MICE PMM generally results in model sizes closest to using the full data set.

\subsubsection{Cut-off value recovery}
\begin{figure}[H]
    \centering
    \includegraphics[width=1.2\linewidth]{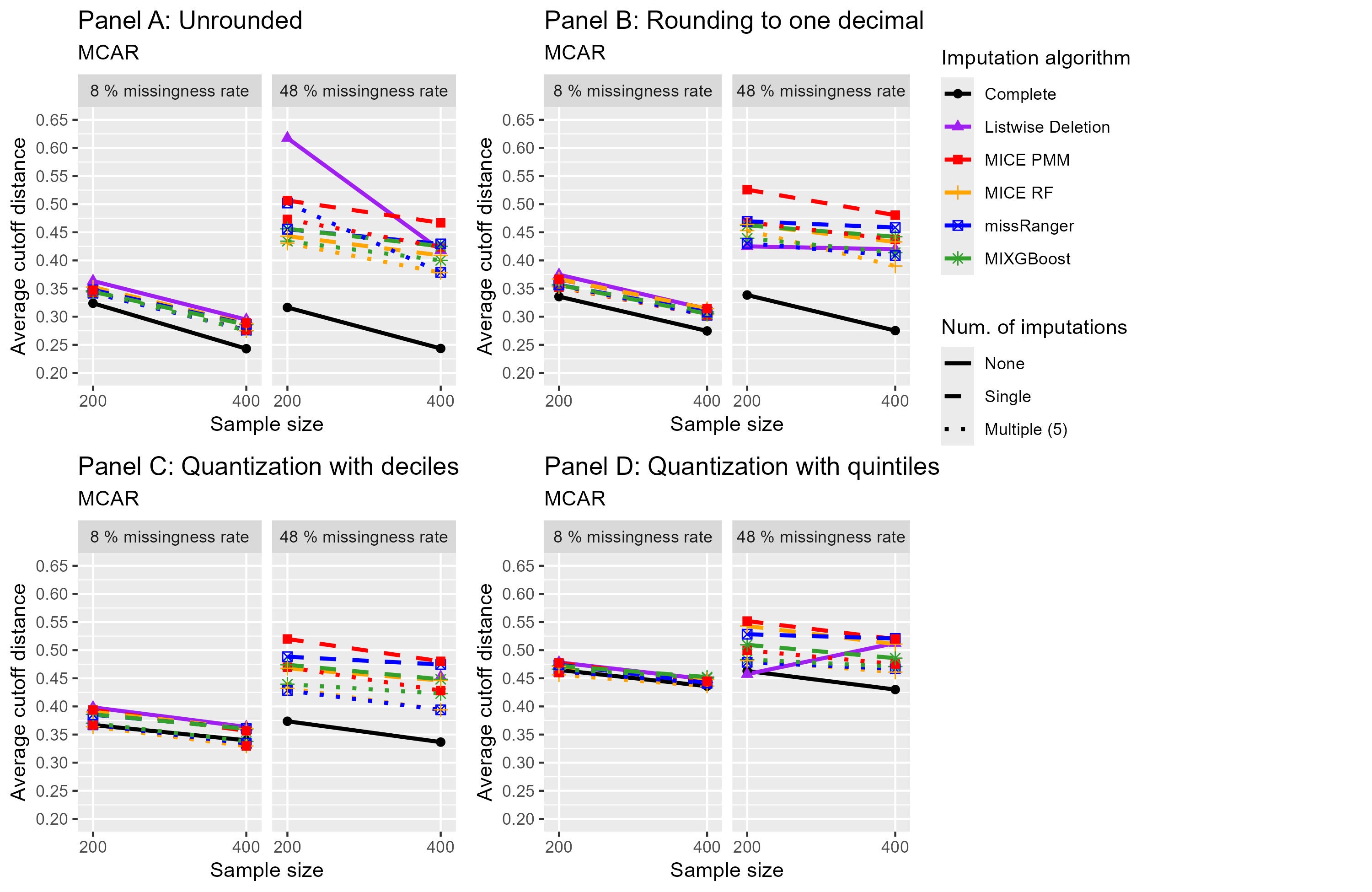}
        \caption{Average cut-off value distance with different imputation methods, under different rates of missingness (MCAR) and different data coarsening settings.}
    \label{fig:CutoffDistance}
\end{figure}

The cut-off value distance, shown in Panel A of Figure~\ref{fig:CutoffDistance}, measures the average absolute distance from the cut-off values used in the estimated rules to the true cut-off values in the data generating rules. In almost all settings, MI models achieve a smaller distance and therefore more accurately estimated rules, with the exception of missRanger when the sample size is small and missingness is large. In most settings, using LD leads to larger distances, except when missingness is high and the sample size is large: In this setting, LD performs better than MICE PMM MI/SI, missRanger SI and MIXGBoost SI. Between the imputation algorithms, MICE RF MI performs the best, e.g., when missingness is high and the sample size is large, using MICE RF MI leads to an average cut-off value distance of 0.431, while the worst performing imputation method, MICE PMM SI, yields an average distance of 0.506. \\[1em]

In summary, without coarsening, MIXGBoost MI achieves the best performance among the imputation algorithms in terms of rule recovery and prediction quality, but has a higher false positive rate and is less interpretable. MI generally performs better than SI with only minor exceptions. Yet, MI results in larger models than SI, so that SI remains an option, especially when computational time is a limiting factor. All imputation methods outperform LD by a large margin, especially when missingness is high, so that LD is generally not advisable to use.

\subsection{Results: Effects of coarsened data} 
\subsubsection{Rule recovery}

In Panel B to D of Figures~\ref{fig:recovery} through~\ref{fig:CutoffDistance}, the differences between the unrounded data (Panel A) and the 3 levels of coarsening (Panel B, C and D) are shown.
A detailed overview of how the different levels of coarsening affect performance with each level of sample size and missingness rate can be seen in Figures~\ref{fig:RRMCARalt} through~\ref{fig:CutoffMCARalt} in the \hyperref[sec:Appendix]{Appendix}.

In terms of rule recovery (Figure~\ref{fig:RRMCARalt} in the \hyperref[sec:Appendix]{Appendix}), rounding to one decimal and quantization to deciles does not lead to any noticeable reduction in performance for any imputation method and in some cases even slightly improves rule recovery. Only quantization to quintiles greatly reduces the ability of the models to recover rules. For example, for MIXGBoost MI on large datasets with low missingness, the rule recovery rate is 0.766 when the data has been left unrounded (Panel A), 0.755 when rounding to one decimal (Panel B), 0.774 when quantization to deciles is applied (Panel C) and 0.730 when using quintiles (Panel D). Similar trends can be observed for the other imputation methods and in different datasets. MIXGBoost MI remains the best performing algorithm in terms of rule recovery for all levels of coarsening and LD remains the worst option.\\

In Table~\ref{tab:RRCoarsening}, rule recovery is shown separately for every rule. Noteworthy is the difference between rules and the linear term. Rule recovery worsens with coarsening for every rule, but for the linear term it appears to improve, from 0.12 to 0.23, averaged over all other settings.

\begin{table}[htpb!]
\caption{Rule recovery for different levels of coarsening.}

\begin{tabular}{llrr rrrrr}
  \toprule
  &&&& \multicolumn{5}{c}{Recovery}\\
    \cmidrule(l){5-9}
    &&&& & \multicolumn{4}{c}{number of Coarsening}\\
    \cmidrule(l){6-9}
  Term & type & coef. & prev. & total & none & 1st & 10\% & 20\% \\ 
  \midrule
  X7 & linear & 0.50 & 1.000 & 0.17 & 0.12 & 0.11 & 0.20 & 0.23 \\ 
  X9 $>$ -1 \& X2 $>$ -1 & rule & 0.75 & 0.702  & 0.13 & 0.15 & 0.15 & 0.12 & 0.10 \\ 
    X10 $>$ 0.5 \& X1 $>$ 0.5 & rule & 1.00 & 0.094  & 0.58 & 0.61 & 0.61 & 0.58 & 0.51 \\ 
    X10 $>$ -0.75 \& X5 $>$ -0.75 & rule & 1.25 & 0.600 & 0.71 & 0.75 & 0.76 & 0.73 & 0.60 \\ 
    X8 $>$ 0.25 \& X1 $>$ 0.25 & rule & 1.50 & 0.161  & 0.51 & 0.52 & 0.53 & 0.51 & 0.47 \\ 
    X6 $>$ -0.5 \& X5 $>$ -0.5 & rule & 1.75 & 0.445 & 0.76 & 0.78 & 0.77 & 0.77 & 0.72 \\ 
    X1 $>$ 0 \& X2 $>$ 0 & rule & 2.00 & 0.299 & 0.86 & 0.86 & 0.86 & 0.86 & 0.85 \\ 
   \bottomrule
   \multicolumn{9}{p{.8\textwidth}}{\emph{Notes:} coef.: true coefficient; prev.: prevalence; total: average over all amounts and conditions; 1st: rounding to first decimal, amounts to a coarsening in steps of 0.1 SD; 10\%: coarsening  to deciles; 20\%: coarsening  to quintiles.}
\end{tabular}
\newline
\label{tab:RRCoarsening}
\end{table}

\subsubsection{Predictive performance and coefficient bias}
As expected, loss of information due to coarsening the data leads to less accurate predictions and to an increase in MSE (Figure~\ref{fig:MSEMCARalt} in the \hyperref[sec:Appendix]{Appendix}). This effect is strongest when the sample size is large and missingness is low (Panel A). In this case, rounding to one decimal leads to a small but noticeable increase in MSE for all imputation methods, LD and for the complete data. When predictions are less accurate to begin with, e.g., when the sample size is small and missingness is high (Panel D), rounding to one decimal does not change the MSE much, except for the complete data set. When using MI, even quantization to deciles does not affect the predictive power much in this case. Single imputation leads to higher MSEs in all data-generating settings and the gap tends to increase when coarsening is introduced. Using quintiles results in a substantial increase in MSE in all cases. Coarsening does not change the order of imputation algorithms in terms of MSE: MIXGBoost achieves the best predictive performance in nearly all data scenarios, followed by MICE PMM, MICE RF and missRanger. LD results in worse predictions compared to all imputation methods with all levels of coarsening.\\
The same trends can be observed for coefficient bias, as seen in Figure~\ref{fig:BiasMCARalt} in the \hyperref[sec:Appendix]{Appendix}, since the MSE is strongly influenced by how accurately the model can estimate the coefficients. Rounding to one decimal and quantization to deciles does not increase the bias much, especially when MI is being used, while quantization to quintiles substantially increases bias in all cases.\\

\subsubsection{False positives and model size}
Coarsening the data noticeably reduces the number of false positives (Figure~\ref{fig:FPMCARalt} in the \hyperref[sec:Appendix]{Appendix}), especially for MIXGBoost and MICE PMM, which tend to result in the largest models.\\ 
This, however, is more pronounced for coarsening to deciles and quintiles. For example, when sample size is large and missingness is high (Panel B), using MICE PMM SI results in 22.896 false positives on average on the unrounded data and in only 15 on average when the data is quantized to deciles. In most cases, rounding to one decimal leads to slight decreases in the number of false positives and quantizing to deciles can lead to a substantial reduction. SI benefits more strongly from coarsening than MI in most settings in terms of the number of false positives. For example, when sample size is large and missingness is high (Panel B), for MICE RF and for missRanger, MI performs better than single imputation when data is unrounded or when rounded to one decimal, while quantizing to deciles reduces the false positives in the SI models more strongly than in the MI models, so that in these cases, SI shows fewer false positives than MI. For MIXGBoost and MICE PMM, MI results in more false positives than SI for all levels of coarsening and all levels of sample size and rate of missingness.\\
The reduction in false positives by coarsening the data directly leads to a reduction in overall model size as well, making the models more interpretable (Figure~\ref{fig:SizeMCARalt} in the \hyperref[sec:Appendix]{Appendix}). MICE RF and missRanger imputed datasets lead to smaller models than MIXGBoost and MICE PMM and SI leads to smaller models than MI.

\subsubsection{Cut-off value recovery}

Coarsening the data lowers the average cut-off value recovery (Figure~\ref{fig:CutoffMCARalt} in the \hyperref[sec:Appendix]{Appendix}), but there are large differences depending on the data setting. When the sample size is large and missingness is low (Panel A), even the  smallest form of coarsening to one decimal already noticeable weakens the cut-off value recovery in each imputation and LD setting. When either the sample size is large and missingness is low (Panel B), or vice versa, the sample size is small but missingness is high (Panel C), coarsening up to deciles does not increase the average cut-off distance by much, especially when MI is used instead of SI. Lastly, when the sample size is small and missingness is high (Panel D), even coarsening to quintiles has only a mild effect on cut-off value recovery.

\subsection{Results: MCAR vs MAR}
When the data missing mechanism is MAR, there is more structure to exploit for the imputation algorithms (see Table~\ref{tab:MARvsMCAR} and Figures~\ref{fig:recoveryMAR} through ~\ref{fig:CutoffDistanceMAR} in the \hyperref[sec:Appendix]{Appendix}). Therefore, as expected, performance generally becomes slightly better under MAR compared to MCAR. When averaged over all settings, rule recovery increases from 0.518 to 0.543, the MSE decreases from 2.32 to 2.26, the coefficient bias decreases from 0.753 to 0.724 and the average cutoff distance decreases from 0.407 to 0.393. However, the number of false positives and the overall model size increase under MAR as well, the false positive rate from 10.7 to 10.9 and the overall model size from 19.8 to 20.6. All observations made in the MCAR case, e.g. the observed differences between the imputation algorithms, the comparison to LD and the differences between MI and SI still hold under MAR. Overall, the missingness mechanism does not seem to interact with the effects of the other varied factors.

\begin{table}[ht]
\centering
\caption{Average result for each performance metric under MCAR and MAR.}
\begin{tabular}{lcccccc}
  \toprule
  & Rule recovery & MSE & Coefficient bias & False positives & Model size & Cut-off distance \\ 
  \midrule
  MCAR & 0.518 & 2.324 & 0.753 & 10.726 & 19.828 & 0.407 \\ 
  MAR & 0.543 & 2.263 & 0.724 & 10.869 & 20.612 & 0.393 \\ 
  \bottomrule
\end{tabular}
\label{tab:MARvsMCAR}
\end{table}

\section{Illustrative example}
In the following, we investigated if the results attained in the simulation hold for an empirical data set as well. For the PREs, the same settings as in the simulation will be used. Most notably, we constrain the tree depth to two and relaxed Lasso is used. To keep things concise, we will only look at five multiple imputations because multiple imputation performed better than single imputation in most cases.

We re-analyse data also analysed by \textcite{SchwerterDBM2022}, who estimate the effect of increased participation in practice tests on exam performance in a first semester gateway math course for economics and business administration students. The data includes, among other things, a survey at the beginning as well as students' performance on three practice tests during the semester.

The PRE models predict performance, the percentage correct score in the last practice test before the final exam, based on a total of 43 predictor variables. These include demographic information, socioeconomic and family background, high school performance, as well as motivational and personality variables including expectancy-value math beliefs \parencite{EcclesEA1983, Gaspard2017}, achievement goals \parencite{ElliotM2008}, big five personality traits \parencite{SchuppG2014}, present-bias preferences \parencite{SchuppG2014}, and self-set course goals.

\begin{table}[htpb!]
\caption{Descriptive Statistics for Sample Variables ($N=325$).}\label{tab:descriptive}
\centering
\scriptsize
\begin{tabular}[t]{p{6.5cm}lrrrrr}
\toprule
\multicolumn{1}{l}{Name } & Abbreviation & \multicolumn{1}{l}{N} & \multicolumn{1}{l}{Mean} & \multicolumn{1}{l}{SD} & \multicolumn{1}{l}{Min} & \multicolumn{1}{l}{Max} \\
\midrule
Performance practice test 3 & Performance & 325 & 0.45 & 0.32 & 0.00 & 1.00\\
Female & Female & 321 & 1.56 & 0.50 & 1.00 & 2.00\\
Income & Income & 243 & 632.69 & 391.97 & 0.00 & 3000.00\\
Rent & Rent & 247 & 299.51 & 211.36 & 0.00 & 1590.00\\
Working & Working & 288 & 0.48 & 0.50 & 0.00 & 1.00\\
Need to work to finance study& Work-study& 295 & 0.25 & 0.43 & 0.00 & 1.00\\
Mother's secondary education background & Mother's sec. edu. back. & 298 & 0.50 & 0.50 & 0.00 & 1.00\\
Father's secondary education background & Father's sec. edu. back. & 295 & 0.54 & 0.50 & 0.00 & 1.00\\
Mother's tertiary education background & Mother's ter. edu. back. & 297 & 2.34 & 0.59 & 1.00 & 3.00\\
Father's tertiary education background & Father's ter. edu. back. & 296 & 2.49 & 0.59 & 1.00 & 3.00\\
Mother's professional status & Mother's occupational status & 300 & 2.89 & 0.99 & 1.00 & 5.00\\
Father's professional status & Father's occ. sta. & 294 & 2.64 & 0.60 & 1.00 & 3.00\\
Highest secondary education of parents & HiSEP & 301 & 0.64 & 0.48 & 0.00 & 1.00\\
Highest tertiary education of parents & HiTEP & 300 & 2.58 & 0.55 & 1.00 & 3.00\\
Highest occupation status of parents & HiOS & 304 & 3.06 & 0.81 & 1.00 & 5.00\\
High school GPA & HS GPA & 315 & 2.19 & 0.62 & 1.00 & 4.00\\
Advanced math in high school & Advanced math & 304 & 0.81 & 0.40 & 0.00 & 1.00\\
Last math grade in high school & Last math grade & 315 & 2.76 & 1.16 & 1.00 & 5.00\\
B. Sc. in international economics & International economics & 325 & 0.44 & 0.50 & 0.00 & 1.00\\
B. Sc. in sport management & Sport management & 325 & 0.05 & 0.22 & 0.00 & 1.00\\
Economics or business administration as a minor & Minor & 325 & 0.17 & 0.38 & 0.00 & 1.00\\
Number of semesters& Semester & 313 & 1.33 & 1.31 & 1.00 & 13.00\\
Repeating the course & Repeater & 313 & 1.99 & 0.21 & 1.00 & 3.00\\
Self-set course goal 1& Goal 1 & 310 & 2.78 & 0.51 & 0.00 & 3.00\\
Self-set course goal 2& Goal 2 & 310 & 0.78 & 0.14 & 0.00 & 1.00\\
Self-set course goal 3& Goal 3 & 310 & 1.26 & 0.46 & 1.00 & 3.00\\
Self-set course goal 4& Goal 4 & 310 & 2.10 & 0.64 & 1.00 & 4.00\\
Self-concept & Self-concept & 318 & 2.65 & 0.64 & 1.25 & 4.00\\
Intrinsic value & Intrinsic value & 316 & 2.69 & 0.61 & 1.00 & 4.00\\
Attainment value & Attainment value & 315 & 2.95 & 0.56 & 1.00 & 4.00\\
Utility value & Utility value & 314 & 3.47 & 0.55 & 1.75 & 4.00\\
Cost & Cost & 315 & 2.39 & 0.55 & 1.00 & 3.83\\
Mastery approach & Mastery approach & 308 & 6.08 & 0.79 & 2.33 & 7.00\\
Mastery avoidance & Mastery avoidance & 308 & 5.60 & 1.01 & 2.67 & 7.00\\
Performance approach & Performance approach & 303 & 4.81 & 1.56 & 1.00 & 7.00\\
Performance avoidance & Performance avoidance & 301 & 4.81 & 1.67 & 1.00 & 7.00\\
Conscientiousness & Conscientiousness & 313 & 4.90 & 1.08 & 0.00 & 7.00\\
Extraversion & Extraversion & 313 & 4.93 & 1.30 & 0.00 & 7.00\\
Agreeableness & Agreeableness & 313 & 5.51 & 1.05 & 0.00 & 7.00\\
Openness & Openness & 311 & 4.91 & 1.17 & 2.00 & 7.00\\
Neuroticism & Neuroticism & 313 & 4.38 & 1.26 & 0.00 & 7.00\\
Risk & Risk & 307 & 0.68 & 0.20 & 0.01 & 1.00\\
Discount factor & Discount factor & 301 & 1.00 & 0.67 & 0.11 & 7.50\\
Present bias & Present bias & 300 & 1.06 & 0.31 & 0.12 & 4.00\\
\bottomrule
\end{tabular}
\end{table}

To evaluate the predictive performance, a simple holdout strategy was employed to test the PRE models on previously unseen data. $N_{train}$ = 217 (66.7$\%$) of the total sample size was used for training the models and the remaining $N_{test} = 108$ observations were used to test them. For a straightforward comparison between imputation methods and listwise deletion, all models were only tested on complete observations from the test set since predictions for data points with missing data would add another layer of complexity. The result is shown in Table~\ref{tab:MSE_Emp}. As expected, LD performs worse than any other imputation algorithm by quite a large margin. Comparing the imputation algorithms, MICE PMM achieves the lowest MSE (0.076), followed by MIXGBoost (0.077), MICE RF (0.078) and missRanger (0.080). Using LD leads to a MSE of 0.097. Overall, the differences between the imputation algorithms are rather minor but follow a similar trend observed in the simulation, with missRanger yielding the least accurate predictions, MICE PMM and MIXGBoost yielding the best predictive performance and MICE RF being in the middle. However, in the simulation, MIXGBoost performed better than MICE PMM, while on the empirical data set, MICE PMM has a small edge on MIXGBoost.\\
The difference between imputation and LD can also be seen in the $R^2$ value for each model, calculated on the hold-out data (Table~\ref{tab:MSE_Emp}). For the LD model, $R^2$ equals to -0.082,which indicates worse predictions than random guessing. MICE PMM yields a $R^2$ of 0.153, followed by MIXGBoost (0.145), MICE RF (0.131) and missRanger (0.103). All imputation models achieve positive $R^2$ values, showing that imputation leads to meaningful prediction models which LD fails to achieve.

\begin{table}[htpb!]
\centering
\caption{MSE, RMSE and $R^2$ evaluated on holdout data for each multiple imputation algorithm and listwise deletion (LD).} 
\begin{tabular}{lccccc}
  \toprule
  Metric & LD & MICE PMM & MICE RF & missRanger & MIXGBoost \\ 
  \midrule
  MSE & 0.097 & 0.076 & 0.078  & 0.080   & 0.077\\ 
  RMSE & 0.311 & 0.276 & 0.279 & 0.284 & 0.277\\
 $R^2$ & -0.082 &  0.153 &  0.131 &  0.103 &  0.145 \\ 
   \bottomrule
\end{tabular}
\label{tab:MSE_Emp}
\newline
{\raggedright\emph{Notes}: $R^2 = 1- \frac{MSE_{test}}{Var(y_{test})}$ }
\end{table}

After the performance evaluation, the models were refit on the entire dataset to mimic a realistic approach to empirical data. Next, the variable importance for the imputation algorithm and for listwise deletion is calculated on the refit models and shown in Table~\ref{tab:importancesEmp} to investigate any possible differences in the variable selection. Most notable is that using LD does not select the variable last math grade in HS, even though it was selected with the third or fourth highest importance when using any of the imputation algorithms. Also, LD selects fathers' occupational status with the joint third highest importance (0.048), while the models trained on imputed data either did not select this variable at all (MICE RF, MIXGBoost) or only with the lowest importance of all the selected variables (MICE PMM, missRanger).

Using imputation, the models always select the five variables HS GPA, part of students minor field of study, extraversion personality trait, last HS math grade and the indicator for whether students need to work to finance their studies with relatively high importances and in very similar orders, with minor differences like MICE PMM selecting HS GPA as the most important variable, while it is only the second most important one for the other algorithms. Compared to LD, the importance of HS GPA is generally somewhat lower. One explanation for this could be that the imputation methods selected a different prior knowledge variable (last math grade in HS). 

Variables not selected consistently by all imputation methods are also the ones with the lowest importance; i.e., important variables are identified by all imputation methods, indicating stability across imputation methods. Examples of low importance variables include the planned number of practice tests (Goal 1; only selected by MICE RF), and the number of semesters of study (Semester; only selected by missRanger).

The relationships between the predictor variables and the dependent variable for each model can be seen in detail in Figure~\ref{fig:univariateDep} in the \hyperref[sec:Appendix]{Appendix}. The direction of the effect, i.e. whether the effect is positive or negative, and the cut-off values of the predictor variable on which the effect changes are very similar across all models, regardless of which imputation method or if listwise deletion was used.

Performance on the third practice test shows a positive correlation with prior (high school) achievement, specifically high school GPA and last math grade. The indicator for advanced math was only included once but still showed a positive correlation. The number of practice tests planned was selected only once as a positive predictor of performance. Negative relations were observed with study-related variables, such as taking the class as a minor and semester of studies. Personality traits showed mixed effects: extraversion was positively correlated with performance, while openness was selected using only some imputation methods and in those cases had a negative correlation. Socioeconomic status showed mixed results, with fathers' occupational status being positively associated (selected only thrice), but income (selected only twice) and working to finance one's studies being negatively associated. Importantly, motivational variables, including expectancy-value beliefs and achievement goals, as well as time preferences, were not selected in any method, indicating no significant incremental effect in predicting practice test performance.

Similar to the results of the simulation, using LD led to the smallest models with only five selected rules. This might be due to the artificial decrease in the sample size by deleting observations with missing values. MICE PMM results in the largest models with 12 selected terms, followed by missRanger, MIXGBoost and MICE RF. This is a difference to the simulation, in which MIXGBoost, on average, yielded the largest models, MICE PMM led to slightly smaller models and MICE RF and missRanger were similar to each other and resulted in much smaller models than MIXGBoost and MICE PMM. Thus, it seems that in empirical data, the differences between the tree-based imputation methods are less major than in the simulation.

\begin{table}[htpb!]
\centering
\footnotesize
\small
\caption{
               Importances of all variables selected at least once after multiple imputation algorithm or listwise deletion (LD). The last row shows the model size for each method.} 
\begin{tabular}{rlrrrrrr}
  \toprule
 &           & \multicolumn{5}{c}{Missing data method}\\
 \cmidrule(lr){3-7}
 & Variable & LD & MICE PMM & MICE RF & missRanger & MIXGBoost & Avg. \\ 
  \midrule
1 & HS GPA & 0.094 & 0.068 & 0.059 & 0.063 & 0.069 & 0.071 \\ 
  2 & Minor & 0.048 & 0.063 & 0.072 & 0.070 & 0.077 & 0.066 \\ 
  3 & Extraversion & 0.050 & 0.042 & 0.031 & 0.036 & 0.032 & 0.039 \\ 
  4 & Last math grade in HS & -- & 0.033 & 0.035 & 0.044 & 0.048 & 0.032 \\ 
  5 & Work to study & 0.012 & 0.017 & 0.019 & 0.015 & 0.023 & 0.017 \\ 
  6 & Fathers' occ. stat. & 0.048 & 0.011 & -- & 0.009 & -- & 0.014 \\ 
  7 & Openness & -- & 0.022 & -- & 0.021 & 0.022 & 0.013 \\ 
  8 & Income & 0.032 & 0.022 & -- & -- & -- & 0.011 \\ 
  9 & Advanced math & -- & 0.014 & -- & 0.016 & -- & 0.006 \\ 
  10 & Goal 1 & -- & -- & 0.016 & -- & -- & 0.003 \\ 
  11 & Semester & -- & -- & -- & 0.015 & -- & 0.003 \\ 
   \midrule
    & Model size & 5 & 12 & 7  & 10 & 8 & 8.400 \\
    \bottomrule
\end{tabular}
\label{tab:importancesEmp}
{\raggedright\emph{Notes}: For variable abbreviations, see Table \ref{tab:descriptive}. Avg.: Average variable importance or model size across all five methods. To see the direction of dependency between the selected variables and the outcome, we included univariate dependency plots in the Appendix, Figure \ref{fig:univariateDep}.}

\end{table}

Finally,  we analysed differences in the combination of variables being selected together in a rule, i.e., their interactions. 
Table~\ref{tab:co-occurrencesAll} shows for all data settings which variables were selected  together in a rule and how often this happened.

Several differences can be observed. For example, the interaction between HS GPA and Minor is contained in every model trained on imputed data, for MICE PMM even twice, but not when using LD. The same holds for the variable combinations HS GPA and last math grade in HS, Minor and Extraversion, and for Minor and last math grade in HS, whose interactions are selected in all models based on imputed data but none based on LD data.

Each model except the one trained on MIXGBoost contains a variable combination which is not present in the other models.
The model trained on LD data is the only one which contains the variables Work-study and HS GPA together in a rule.  The variables Extraversion and Openness appear together twice when using MICE PMM, but never together in any other model. However, all other interactions selected using MICE PMM were also selected by every other model.
All models contain interactions between HS GPA and several other variables, but only using MICE RF leads to the interaction between  HS GPA and Goal 1.
The variables HS GPA and Advanced math only co-occur after imputation with missRanger. Selection of variable interactions is less stable across imputation methods than variable selection in general. 
Figures~\ref{fig:bivDepLD},\dots,\ref{fig:bivDepMix} in the \hyperref[sec:Appendix]{Appendix} show the bivariate dependencies for all rules selected by each model. Overall, variables interact very similarly with each other across all models, if they have been selected together in a rule. The direction of the interaction, i.e. whether the increase or decrease of both variables together has a positive or negative effect on practice test performance, is always the same, and the selected cut-off values are very close to each other as well.

\begin{table}[H]
\centering
\footnotesize
\caption{Interactions of variables for each missing data method.} 
\begin{tabular}{lllccccc}
  \toprule
 & &   & \multicolumn{5}{c}{Missing data method} \\  \cmidrule(lr){4-8} 
 \multicolumn{3}{c}{Pair of variables}   & LD & MICE PMM & MICE RF & missRanger & MIXGBoost \\ 
  \midrule
HS GPA &\&& Extraversion & 2 & 1  & 1 & 2 & 1  \\ 
HS GPA &\&& Minor & 0 & 2 & 1 & 1 & 1 \\
HS GPA &\&& Last math grade in HS & 0 & 1 & 1 & 1 & 2 \\

HS GPA &\&& Income & 1 & 1 & 0 & 0 & 0  \\ 
HS GPA &\&& Work to study & 1 & 0 & 0 & 0 & 0  \\ 
HS GPA &\&& Income & 1 & 1 & 0 & 0 & 0 \\
HS GPA &\&& Selfset course goal 1 & 0 & 0 & 1 & 0 & 0 \\
HS GPA &\&& Advanced math in HS & 0 & 0 & 0 & 1 & 0 \\
Minor &\&& Extraversion & 0 & 1 & 1 & 1 & 1 \\
Minor &\&& Father's occ. stat. & 1 & 1 & 0 & 1 & 0  \\ 
Minor &\&& Last HS math grade & 0 & 1 & 1 & 1 & 1 \\
Minor &\&& Work to study & 0 & 1 & 1 & 0 & 1 \\
Extraversion &\&& Openness & 0 & 2 & 0 & 0 & 0 \\
Work to study &\&& Semester & 0 & 0 & 0 & 1 & 0 \\

   \bottomrule
\end{tabular}
\label{tab:co-occurrencesAll}
\end{table}

Overall, the analysis of the empirical dataset confirms that listwise deletion leads to a large decrease in predictive performance. Using MICE PMM yields the smallest MSE, followed by MIXGBoost, MICE RF and missRanger. Additionally, using LD leads to some differences in variable selection compared to using imputation. Among the imputation algorithms, there are few differences for variables that have an overall high importance, but noticeable differences for those variables with low importances exist. 
The main effects for selected variables and their interaction effects are very similar across all models, conditioned on whether they have been selected at all by a model.
There is limited consistency in the selection of variable interactions, showing that capturing interactions in tree-based methods is a difficult task \parencite{RangerWright2017}. In the absence of a ground truth, as in the simulation, it is difficult to say which algorithm captured the true underlying mechanisms the best.

\section{Discussion}
We conclude that PREs can be fitted to multiple imputed data in both MCAR and MAR scenarios, especially with the use of stacking. Different imputation methods will provide a different trade-off in terms of the number of selected rules and predictive performance. Although running PRE after imputation with MIXGBoost has the best rule recovery and lowest MSE, it also has a higher number of false positives and thus a larger model size. In contrast, missRanger and MICE RF provide lower rule recovery but also fewer false positives, resulting in smaller (and easier to interpret) models. 
Single imputation yields larger MSE and mostly weaker rule recovery than MI, but fewer false positives and smaller models.
Listwise deletion shows substantially worse predictive performance and rule recovery compared to all MI and SI imputation methods, both in the simulation and the illustrative example. The gap tends to increase as missingness increases and sample size decreases in the simulation. 

In the illustrative example, variable selection and variable importance are quite similar between imputation methods, especially for variables with relatively high importances. Variable selection can differ more strongly for variables with low importances and for variable interactions.

Rounding variables to the first decimal makes rules easier to interpret and display without sacrificing performance. Quantizing to deciles improves interpretability by reducing false positives and overall model size, but considerably increases MSE. Quantizing to quintiles leads to a large decrease in performance in all settings.

\subsection{Recommendations}
Based on our simulations, we provide several actionable recommendations to guide empirical researchers when applying PRE with missing data. First, our results suggest that imputation should generally be preferred over listwise deletion (LD). Second, in the single versus multiple imputation debate, multiple imputation shows superior performance in terms of rule recovery and predictive accuracy compared to single imputation, although single imputation may sometimes be preferred because it yields less complex ensembles and requires less computational effort. 

Third, in choosing an imputation method, several trade-offs should be taken into account: Among the imputation methods evaluated, MIXGBoost excels in rule recovery and MSE, but tends to produce more false positives and larger models. Therefore, it is best suited when the primary goal is outcome prediction. On the other hand, MICE RF and missRanger are preferable when the aim is to minimize false positives and maintain smaller model sizes. MICE PMM provides a balanced option, although its performance is closer to that of MIXGBoost.

Fourth, in terms of data processing, rounding to one decimal place (equivalent to 0.1 standard deviation) is a safe practice that generally does not significantly alter the data, while potentially leading to more streamlined models. Quantizing to deciles may reduce false positives but may also reduce predictive power, with these effects being more pronounced for SI than for MI. Quintile quantization did not prove beneficial in our analysis and is therefore not recommended.

Finally, a more general recommendation: The distinction between Missing At Random (MAR) and Missing Completely At Random (MCAR) appeared marginal in our study. This seems to suggest that the distinction should not significantly influence decision making. In particular, MAR can improve the effectiveness of imputation if variables predictive of missingness are available and included in the analyses. However, we caution against assuming that M(C)AR always holds: If the missing data mechanism is MNAR, none of the imputation strategies discussed here is valid.

\subsection{Limitations}
As in all simulation studies, we rely on synthetic data with a limited number of scenarios. Because of the computational burden, we limited sample size to $N=400$.  Larger datasets challenge the capabilities of current PRE imputations, especially with stacked multiple imputed data sets. In this case, reducing the sample fraction of randomly selected training observations used to produce each tree lowers the computational cost considerably.  We expect the trade-off in terms of predictive accuracy to be low when tree depth is limited to 2 or 3. Similarly, opting for single imputation may be a practical alternative for larger datasets, with the disadvantages compared to MI discussed above. 

Empirically, tree ensembles like PREs and the Random Forest often outperform linear model approaches in terms of predictive accuracy, which is often attributed to the capabilities of tree ensembles to capture non-linear relationships and interactions. However, the absence of a specific interaction in the model or a relatively low importance of an interaction effect should not be misinterpreted: \textcite{wright2016little} show that main effects can mask interactions in tree ensembles.  We advise using an approach like that of \textcite{hornung2022interaction} when interactions are of substantial interest.

\subsection{Outlook}
PREs are a mature, interpretable statistical learning method. When study design and a priori variable choice are informed by multiple theories, PREs are able to handle a large number of predictors. An integration of competing or complementary theories becomes possible, and (aggregates of) feature importance inform the relative predictive value of several theories.

Despite our demonstration that moderate coarsening is actually beneficial in terms of model size, other factors also increase model size: Especially, PREs become unwieldy when there are a large number of predictors. Future work will have to investigate decreases in model size without loss of predictive performance, e.g., in situations with highly correlated predictors.  However, it is unclear whether dimensionality reduction techniques like principal component analysis keep non-linear interactions on the level of the original variables intact.

\section{Open Practice}

Materials and analysis code for the simulation study are available at \href{https://osf.io/hjvb7/?view_only=b9f9ec2c37b246359f7ff699c95a76bd}{https://osf.io/hjvb7/?view\_only=b9f9ec2c37b246359f7ff699c95a76bd}. None of the reported studies were preregistered. Due to data protection rules, the data for the illustrative example cannot be made available, but documented code for the analyses in this article is available at \href{https://osf.io/hjvb7/?view_only=b9f9ec2c37b246359f7ff699c95a76bd}{https://osf.io/hjvb7/?view\_only=b9f9ec2c37b246359f7ff699c95a76bd} and is illustrated with simulated data.

\printbibliography
\newpage
\appendix
\renewcommand{\thefigure}{A.\arabic{figure}}
\renewcommand{\thetable}{A.\arabic{table}}
\label{sec:Appendix}

\begin{table}[H]
\caption{Rule Recovery for complete data, listwise deletion, single and multiple imputation, averaged over all imputation algorithms.}
\begin{tabular}{llrr rrrrr}
  \toprule
  &&&& \multicolumn{5}{c}{Recovery}\\
    \cmidrule(l){5-9}
    &&&& & \multicolumn{4}{c}{Missing data method}\\
    \cmidrule(l){6-9}
  name & type & coef. & prev. & total & complete & LD & MI & SI \\ 
  \midrule
X7 & linear Term & 0.50 & 1.000 & 0.17 & 0.26 & 0.18 & 0.19 & 0.13 \\ 
   X9 $>$ -1 \& X2 $>$ -1 & rule & 0.75 & 0.702  & 0.13 & 0.21 & 0.06 & 0.13 & 0.14 \\ 
  X10 $>$ 0.5 \& X1 $>$ 0.5 & rule & 1.00 & 0.094  & 0.58 & 0.63 & 0.27 & 0.58 & 0.58 \\ 
  X10 $>$ -0.75 \& X5 $>$ -0.75 & rule & 1.25 & 0.600  & 0.71 & 0.87 & 0.36 & 0.71 & 0.72 \\ 
  X8 $>$ 0.25 \& X1 $>$ 0.25 & rule & 1.50 & 0.161  & 0.51 & 0.81 & 0.30 & 0.51 & 0.50 \\ 
  X6 $>$ -0.5 \& X5 $>$ -0.5 & rule & 1.75 & 0.445  & 0.76 & 0.99 & 0.51 & 0.75 & 0.78 \\ 
   X1 $>$ 0 \& X2 $>$ 0 & rule & 2.00 & 0.299 & 0.86 & 1.00 & 0.56 & 0.84 & 0.88 \\ 
   \bottomrule
\end{tabular}
\label{tab:RR_LDvsImp}
\newline
{\emph{Notes:} coef.: coefficients; prev.: prevalence; total: average over all variants and conditions; LD: listwise deletion; MI: multiple imputation; SI: single imputation.}

\end{table}

\subsection{Detailed MAR results}

\begin{figure}[H]
    \centering

    \includegraphics[width=1.2\linewidth]{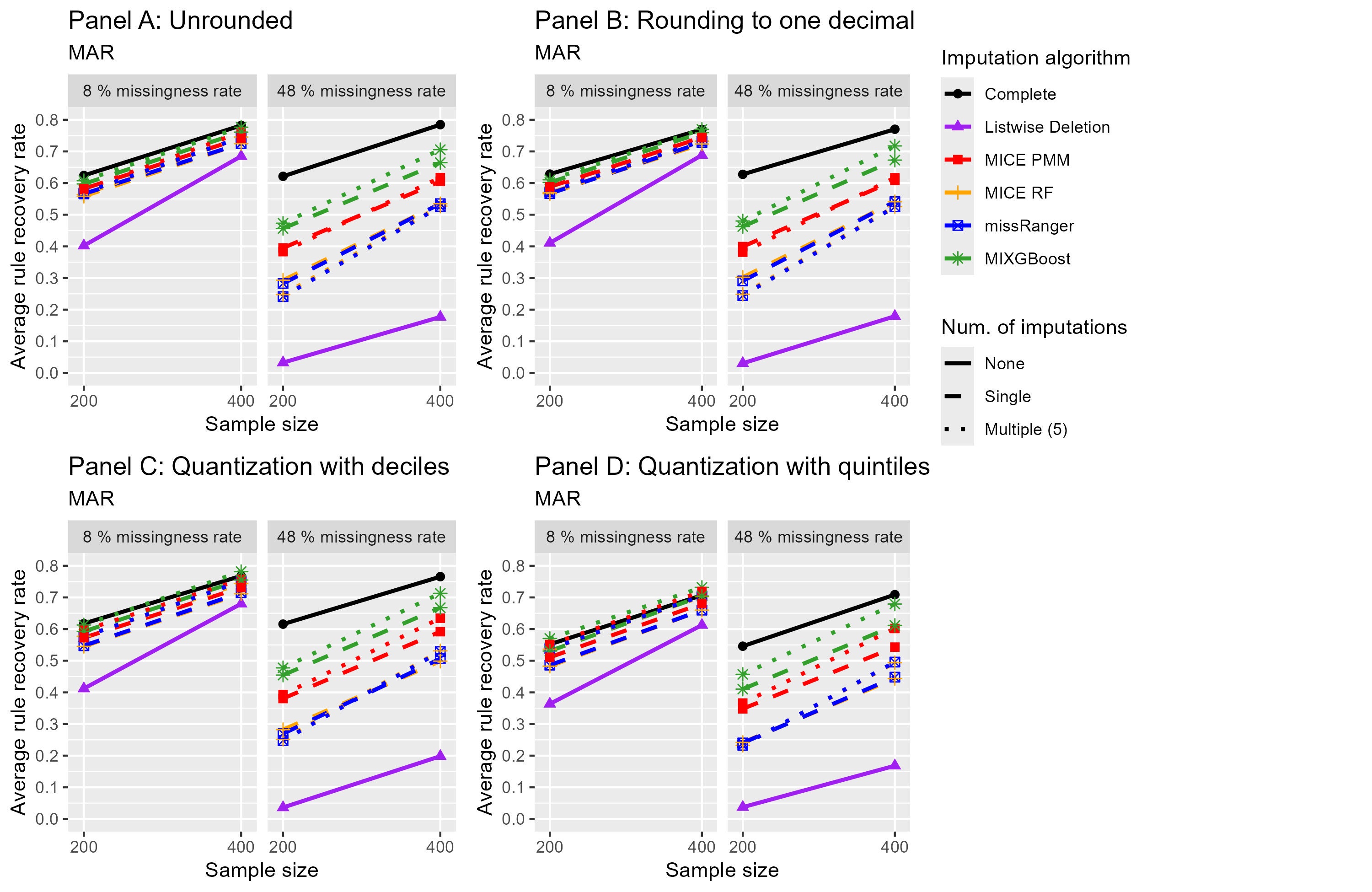 }
        \caption{Rule recovery with different imputation methods, under different rates of missingness (MAR) and different data coarsening settings.}
    \label{fig:recoveryMAR}
\end{figure}

\begin{figure}[H]
    \centering
    \includegraphics[width=1.2\linewidth]{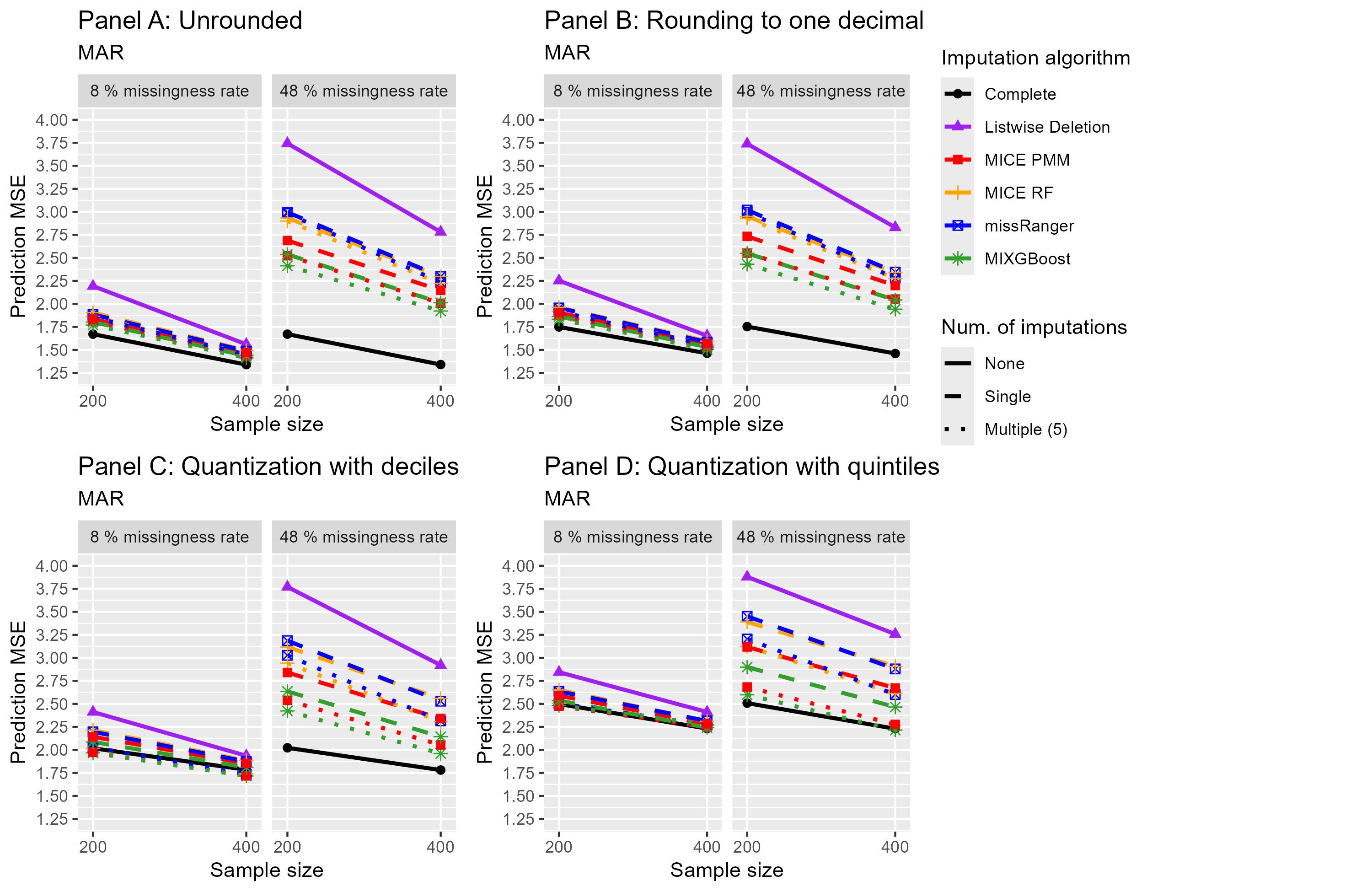 }
    \caption{Predictive performance evaluated by the mean squared error (MSE) with different imputation methods, under different rates of missingness (MAR) and different data coarsening settings.}
    \label{fig:MSEMAR}
\end{figure}

\begin{figure}[H]
    \centering
    \includegraphics[width=1.2\linewidth]{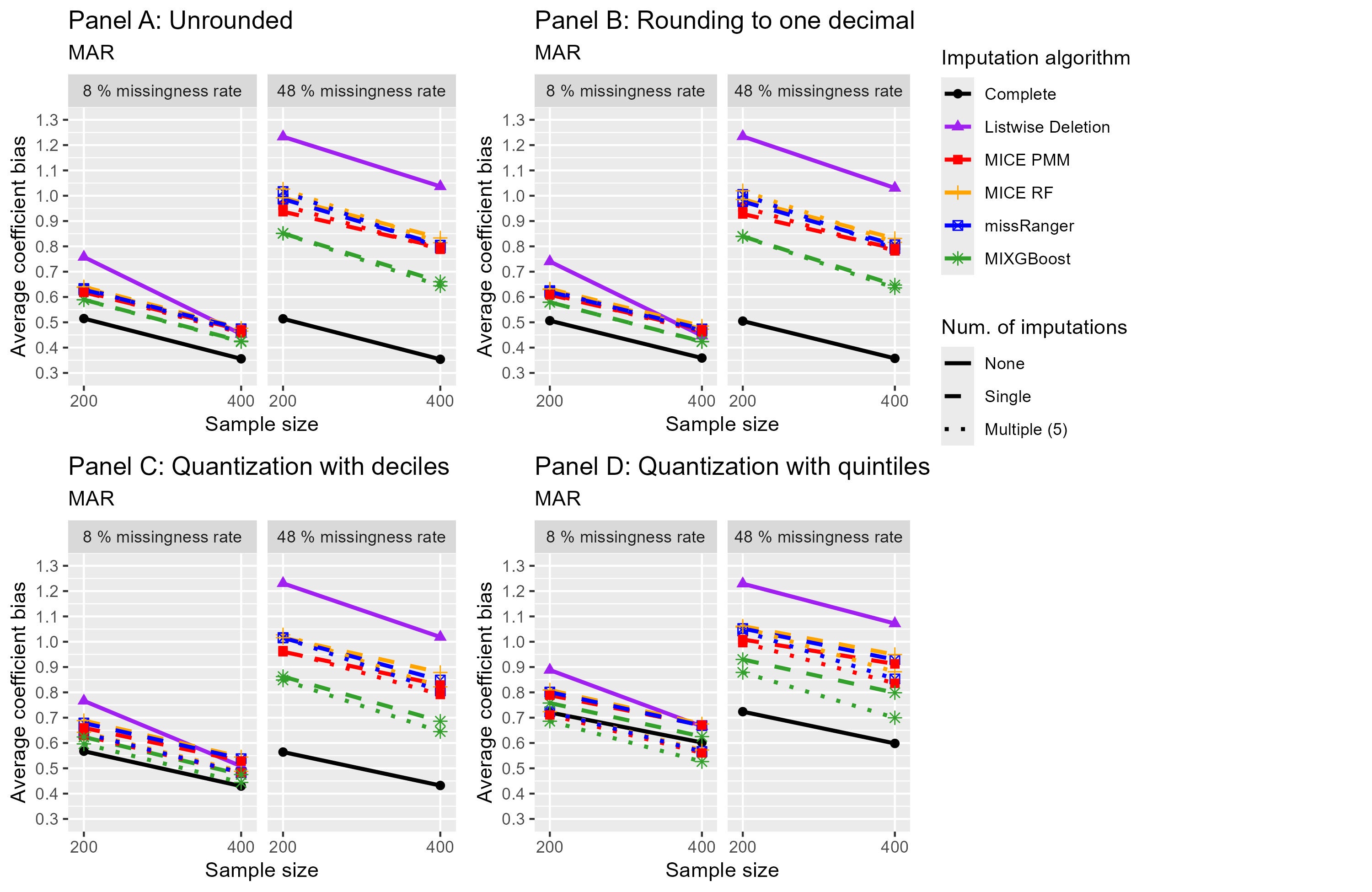}
    \caption{Average coefficient biases with different imputation methods, under different rates of missingness (MAR) and under different data coarsening settings.}
    \label{fig:biasMAR}
\end{figure}

\begin{figure}[H]
    \centering
    \includegraphics[width=1.2\linewidth]{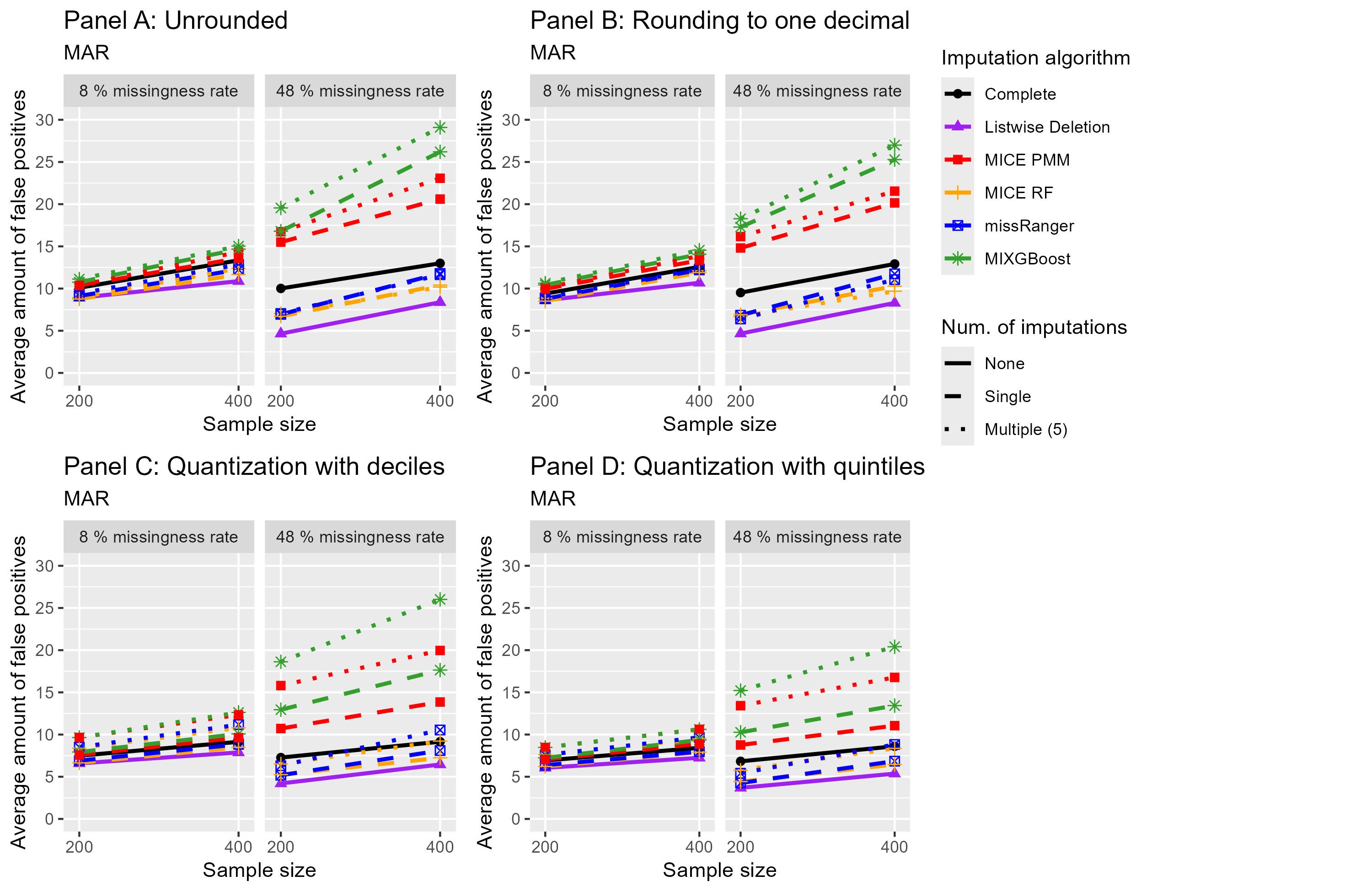 }
    \caption{Average number of false positives with different imputation methods, under different rates of missingness (MAR) and different data coarsening settings.}
    \label{fig:FPMAR}
\end{figure}

\begin{figure}[H]
    \centering
    \includegraphics[width=1.2\linewidth]{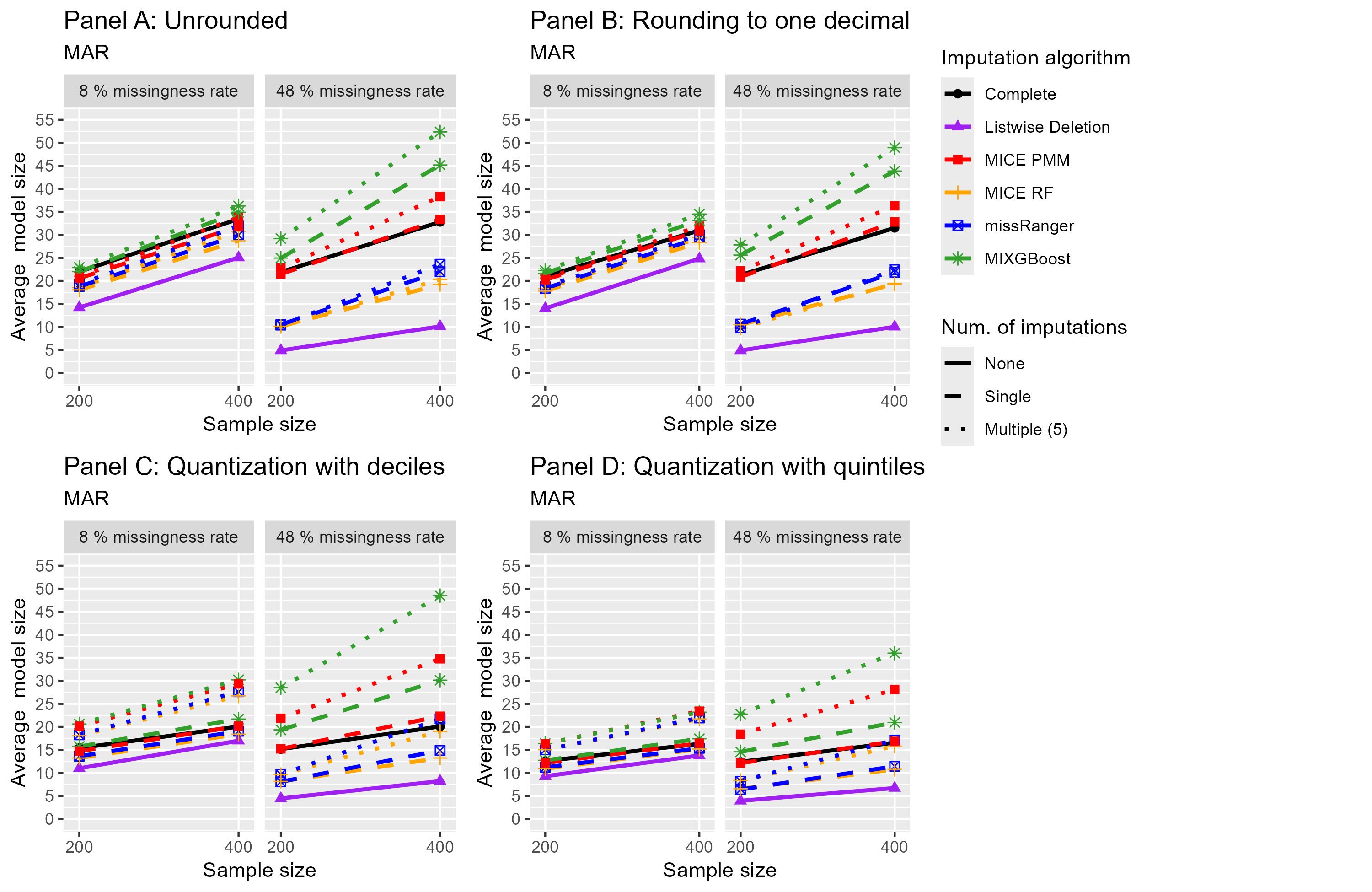}
    \caption{Average model size with different imputation methods, under different rates of missingness (MAR) and different data coarsening settings.}
    \label{fig:SizeMAR}
\end{figure}

\begin{figure}[H]
    \centering
    \includegraphics[width=1.2\linewidth]{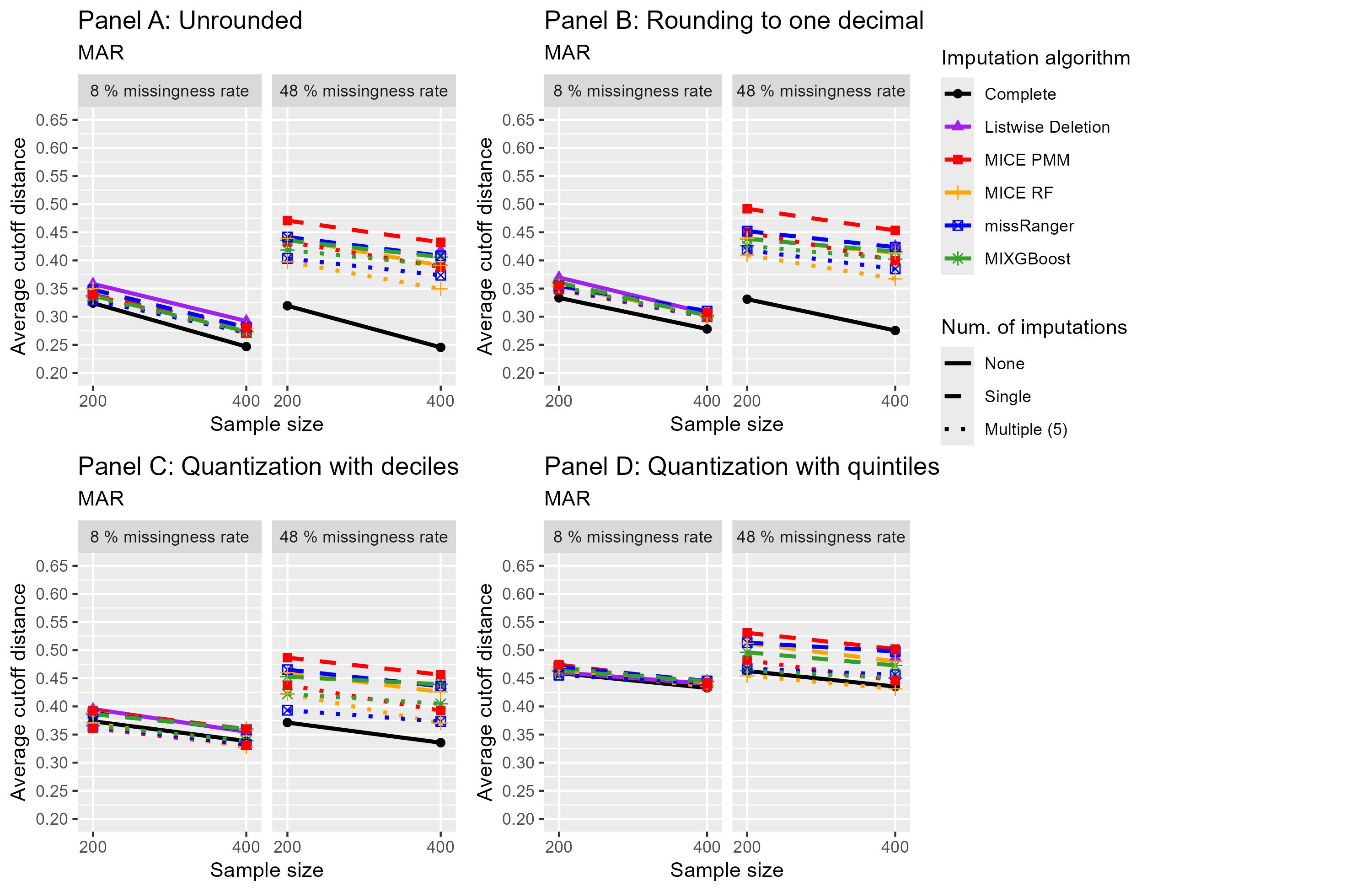}
    \caption{Average Cut-off value distance with different imputation methods, under different rates of missingness (MAR) and different data coarsening settings.}
    \label{fig:CutoffDistanceMAR}
\end{figure}


\subsection{Detailed results under coarsening}
\begin{figure}[H]
    \centering
    \includegraphics[width=1\linewidth]{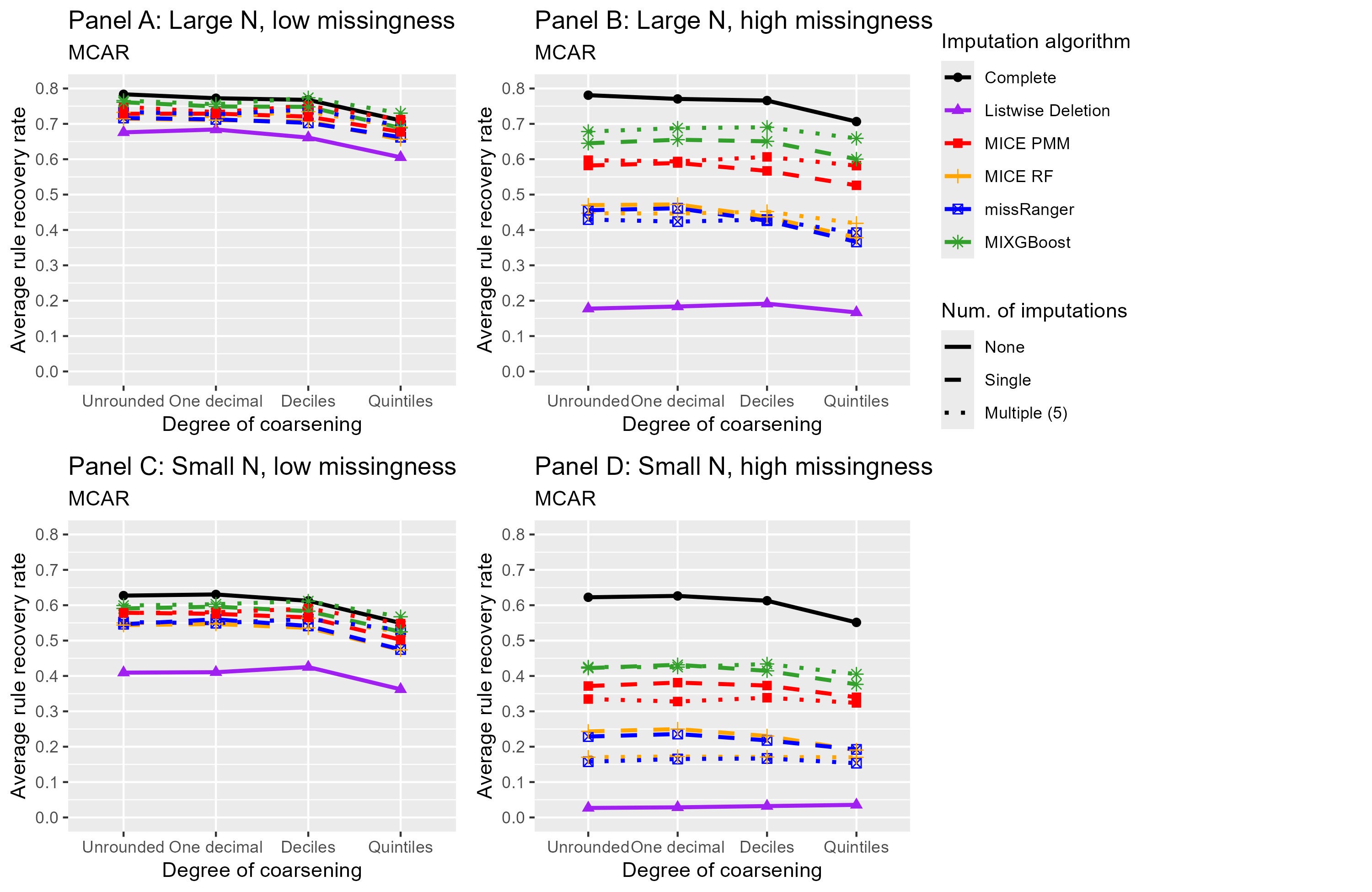}
        \caption{Average rule recovery with different imputation methods, under different data coarsening settings, different sample sizes and different missingness rates (MCAR).}
    \label{fig:RRMCARalt}
\end{figure}

\begin{figure}[H]
    \centering
    \includegraphics[width=1\linewidth]{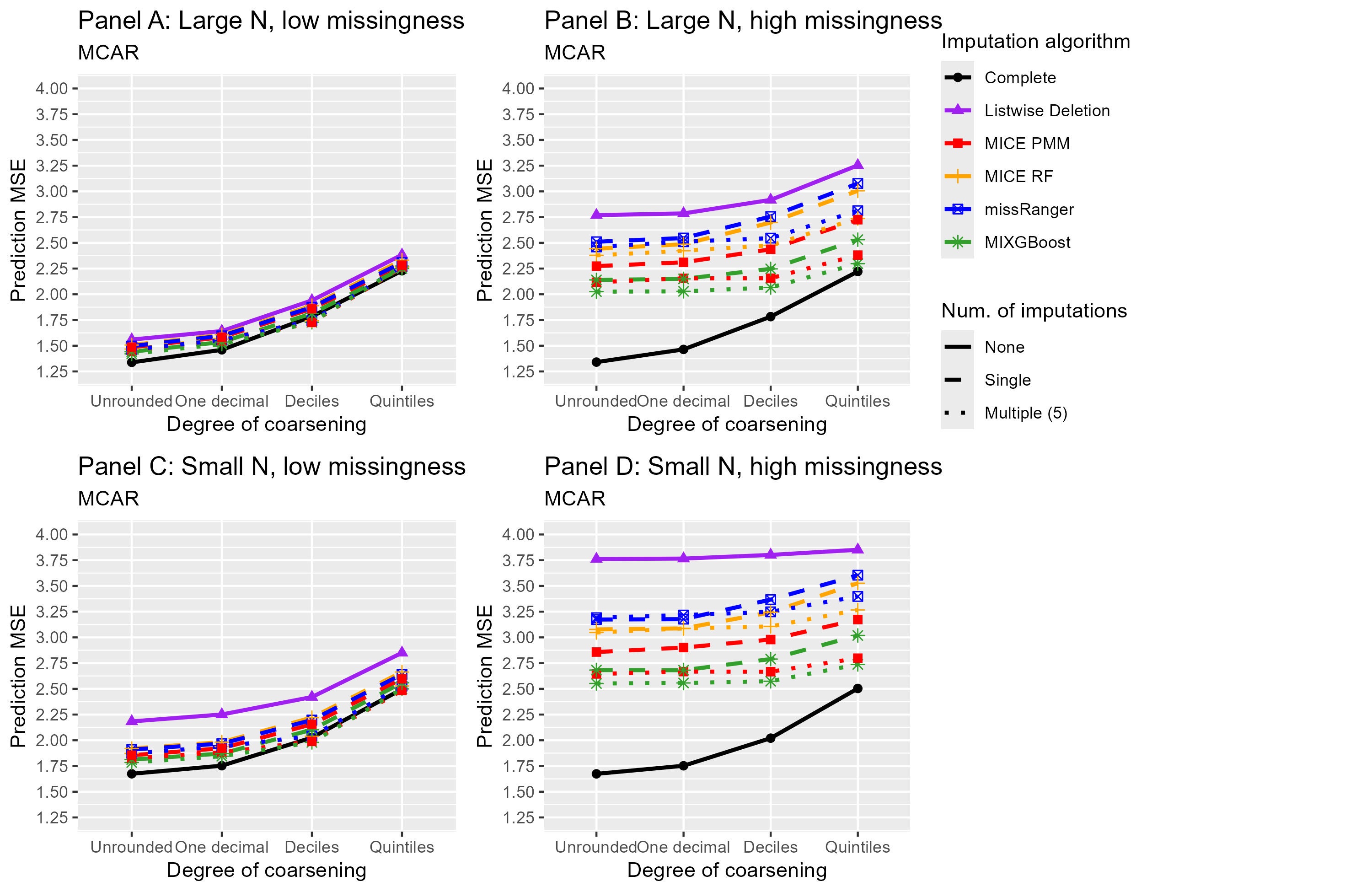}
        \caption{Average prediction MSE with different imputation methods, under different data coarsening settings, different sample sizes and different missingness rates (MCAR).}
    \label{fig:MSEMCARalt}
\end{figure}

\begin{figure}[H]
    \centering
    \includegraphics[width=1\linewidth]{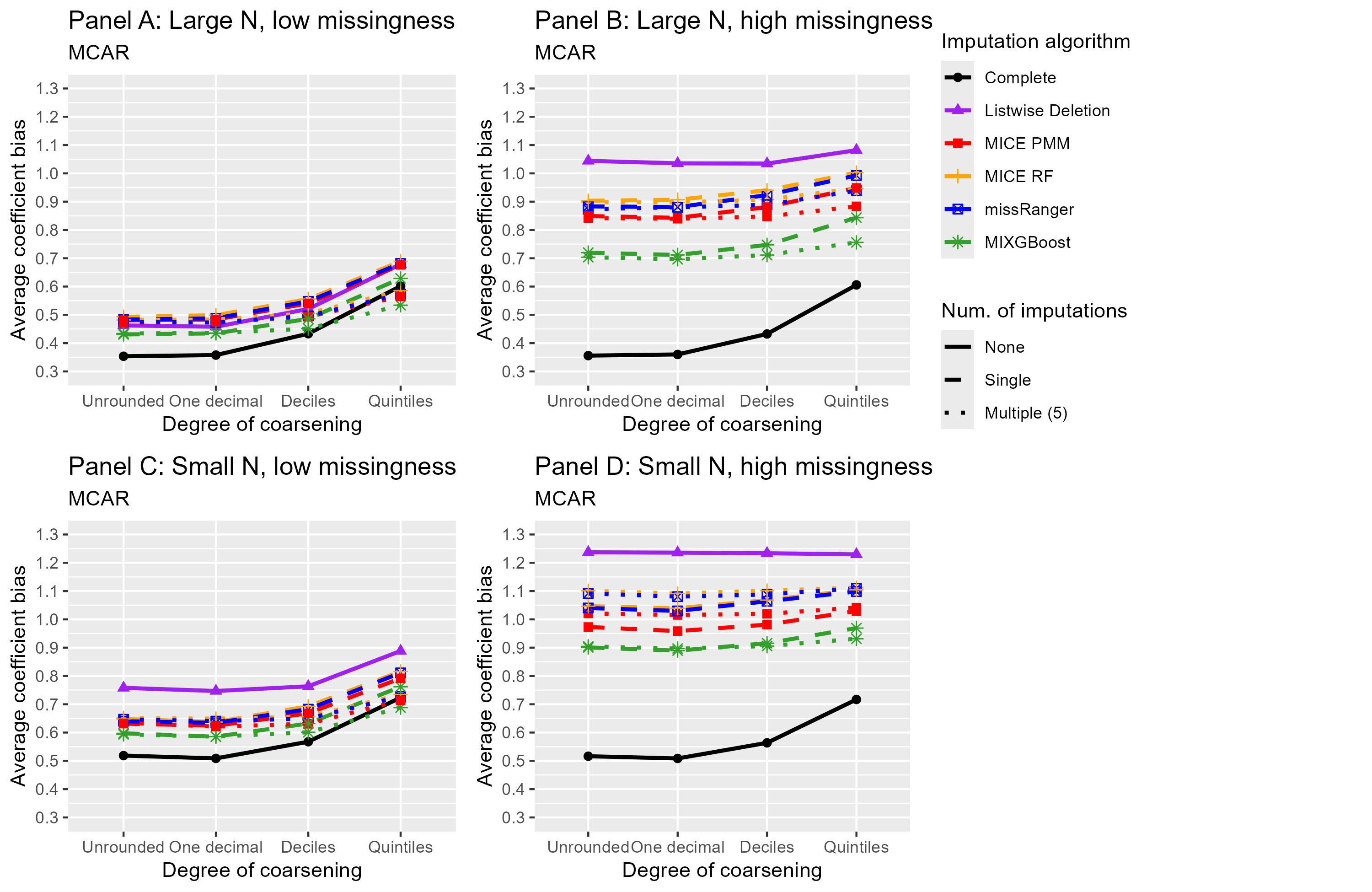}
        \caption{Average coefficient bias with different imputation methods, under different data coarsening settings, different sample sizes and different missingness rates (MCAR).}
    \label{fig:BiasMCARalt}
\end{figure}

\begin{figure}[H]
    \centering
    \includegraphics[width=1\linewidth]{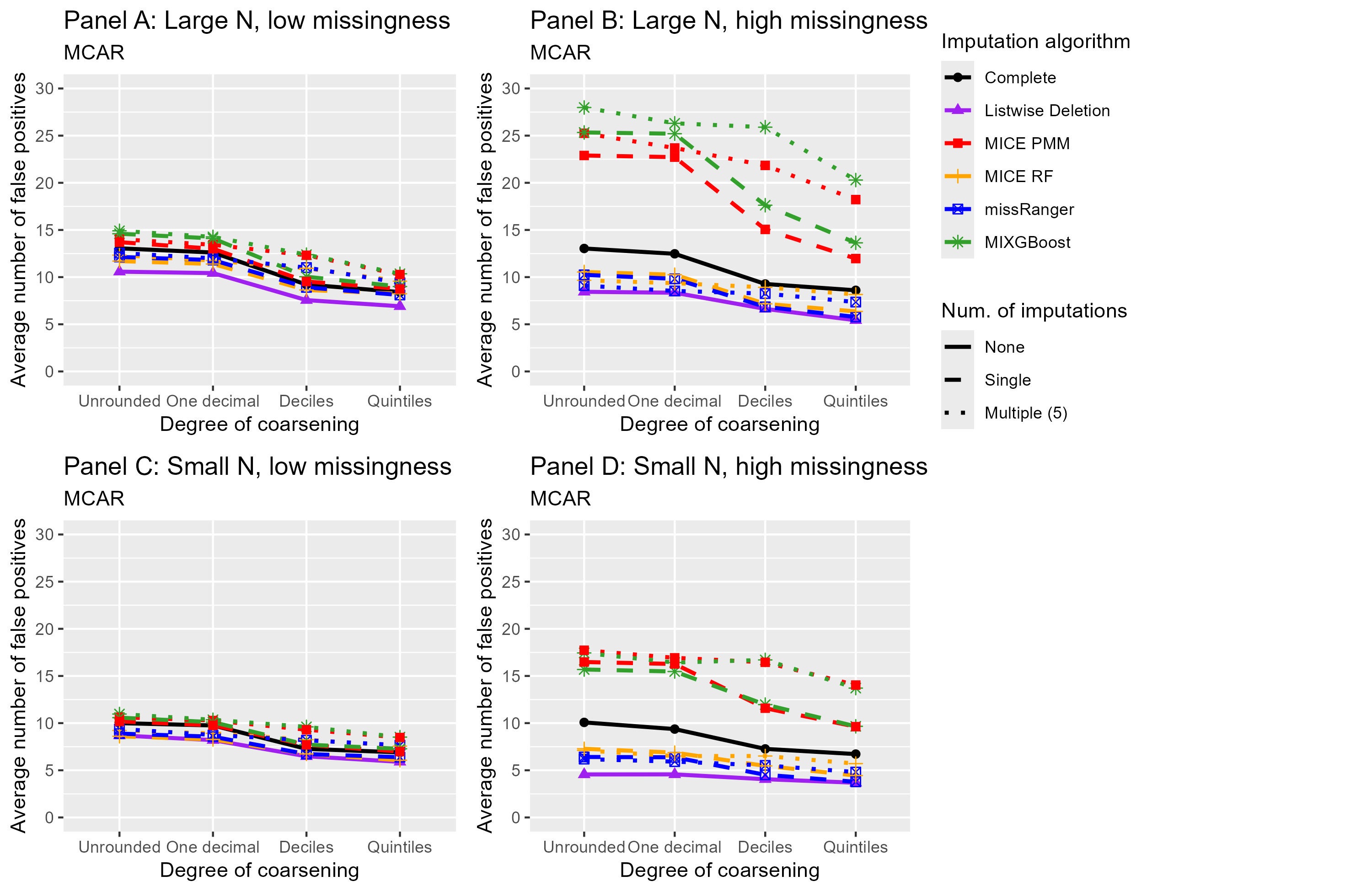 }
        \caption{Average number of false positives with different imputation methods, under different data coarsening settings, different sample sizes and different missingness rates (MCAR).}
    \label{fig:FPMCARalt}
\end{figure}

\begin{figure}[H]
    \centering
    \includegraphics[width=1\linewidth]{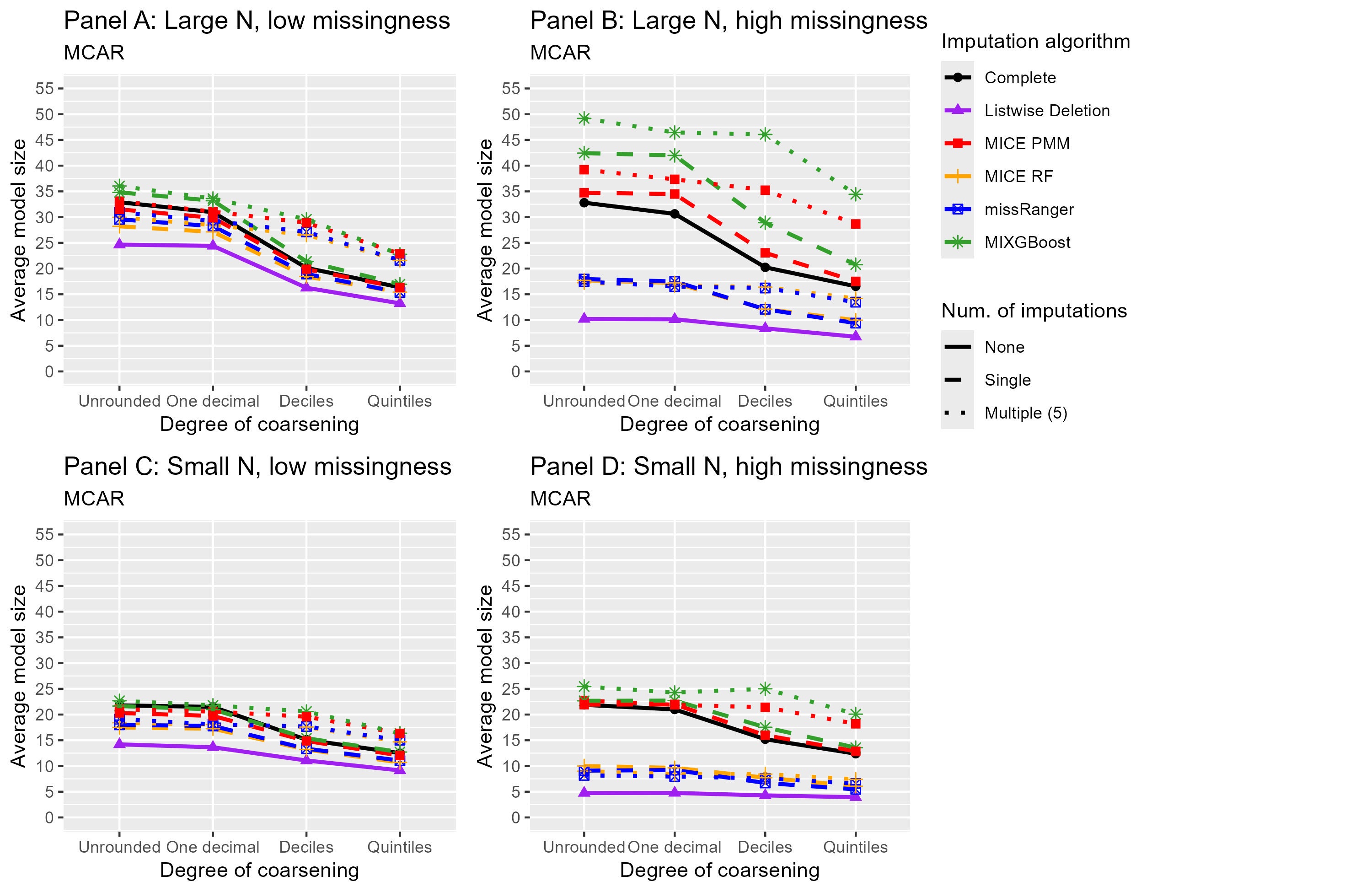}
        \caption{Average model size with different imputation methods, under different data coarsening settings, different sample sizes and different missingness rates (MCAR).}
    \label{fig:SizeMCARalt}
\end{figure}

\begin{figure}[H]
    \centering
    \includegraphics[width=1\linewidth]{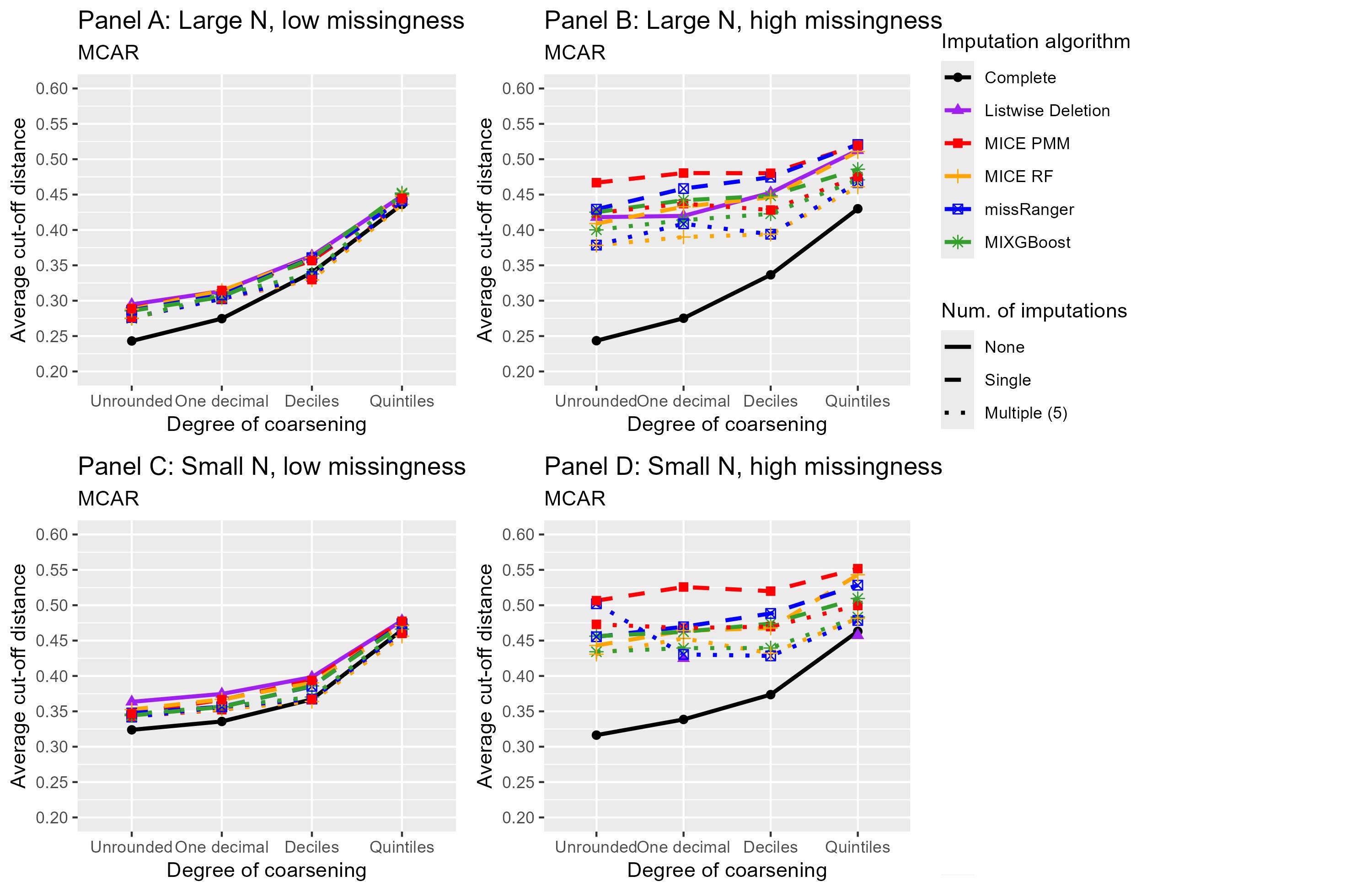}
        \caption{Average cut-off distance with different imputation methods, under different data coarsening settings, different sample sizes and different missingness rates (MCAR).}
    \label{fig:CutoffMCARalt}
\end{figure}

\subsection{Detailed results in the illustrative example}
\begin{figure}[H]
    \centering
\caption{Univariate dependencies in the illustrative example for each model and selected variables on practice test performance.}
    \label{fig:univariateDep}    
    \includegraphics[width=0.9\linewidth]{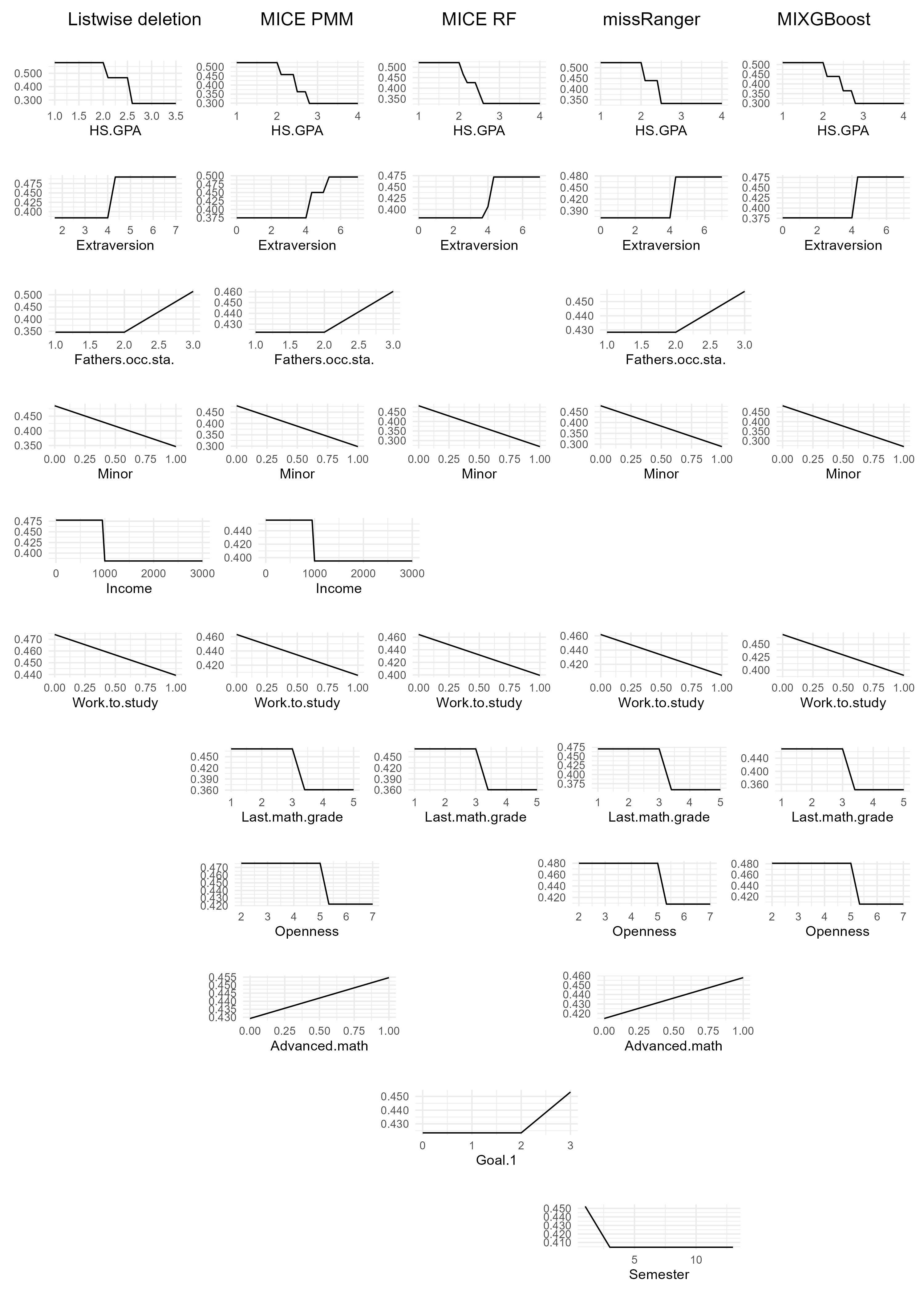}

\end{figure}

\begin{figure}[H]
    \centering
\caption{Bivariate dependencies for all rules selected on practice test performance in the illustrative example for the model trained after using listwise deletion.} 
    \includegraphics[width=1\linewidth]{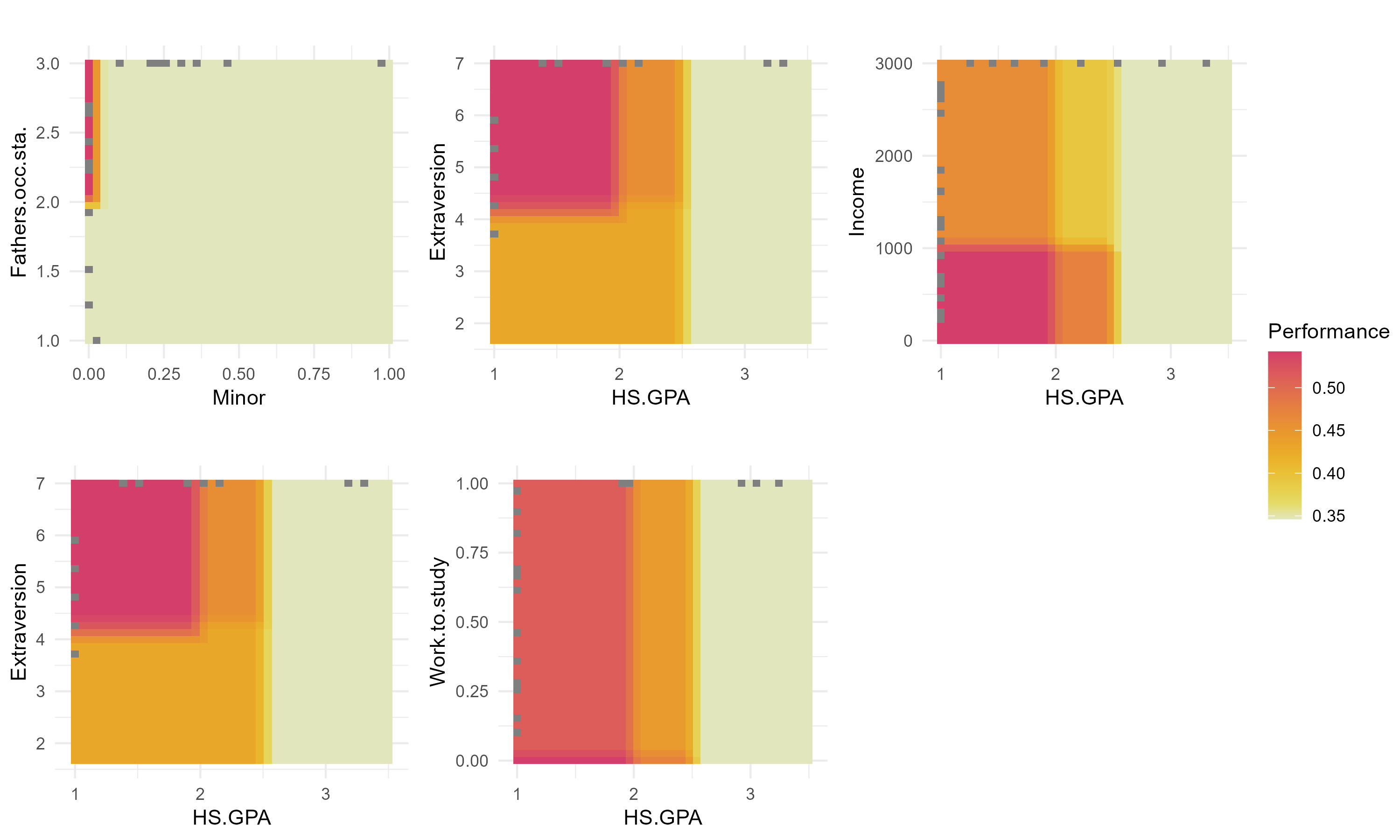}
    \label{fig:bivDepLD}
\end{figure}

\begin{figure}[H]
    \centering
\caption{Bivariate dependencies for all rules selected on practice test performance in the illustrative example for the model trained after using MICE PMM.}
    \label{fig:bivDepPMM}    
    \includegraphics[width=1\linewidth]{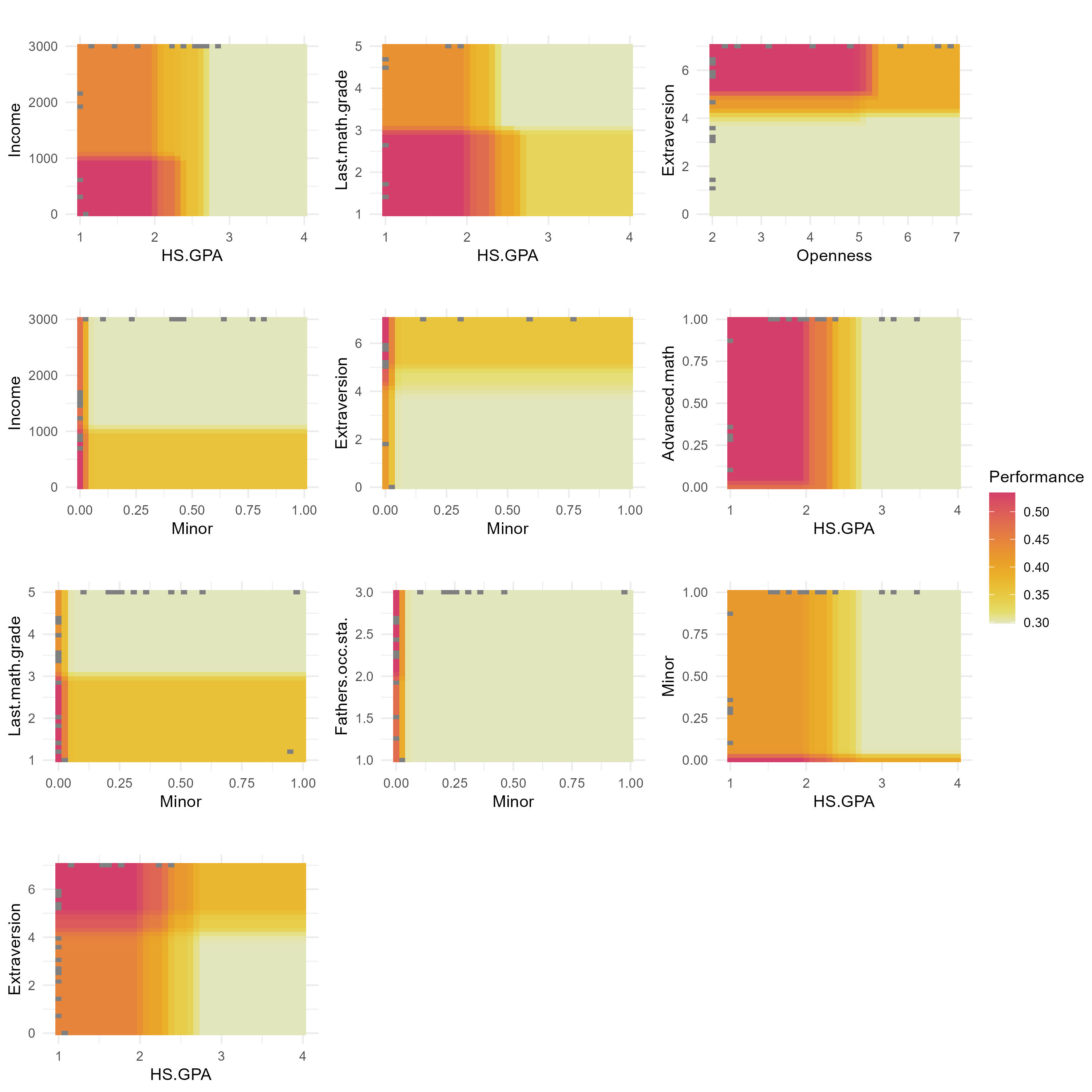}
\end{figure}

\begin{figure}[H]
    \centering
\caption{Bivariate dependencies for all rules selected on practice test performance in the illustrative example for the model trained after using MICE RF.}
    \label{fig:bivDepRF}    
    \includegraphics[width=1\linewidth]{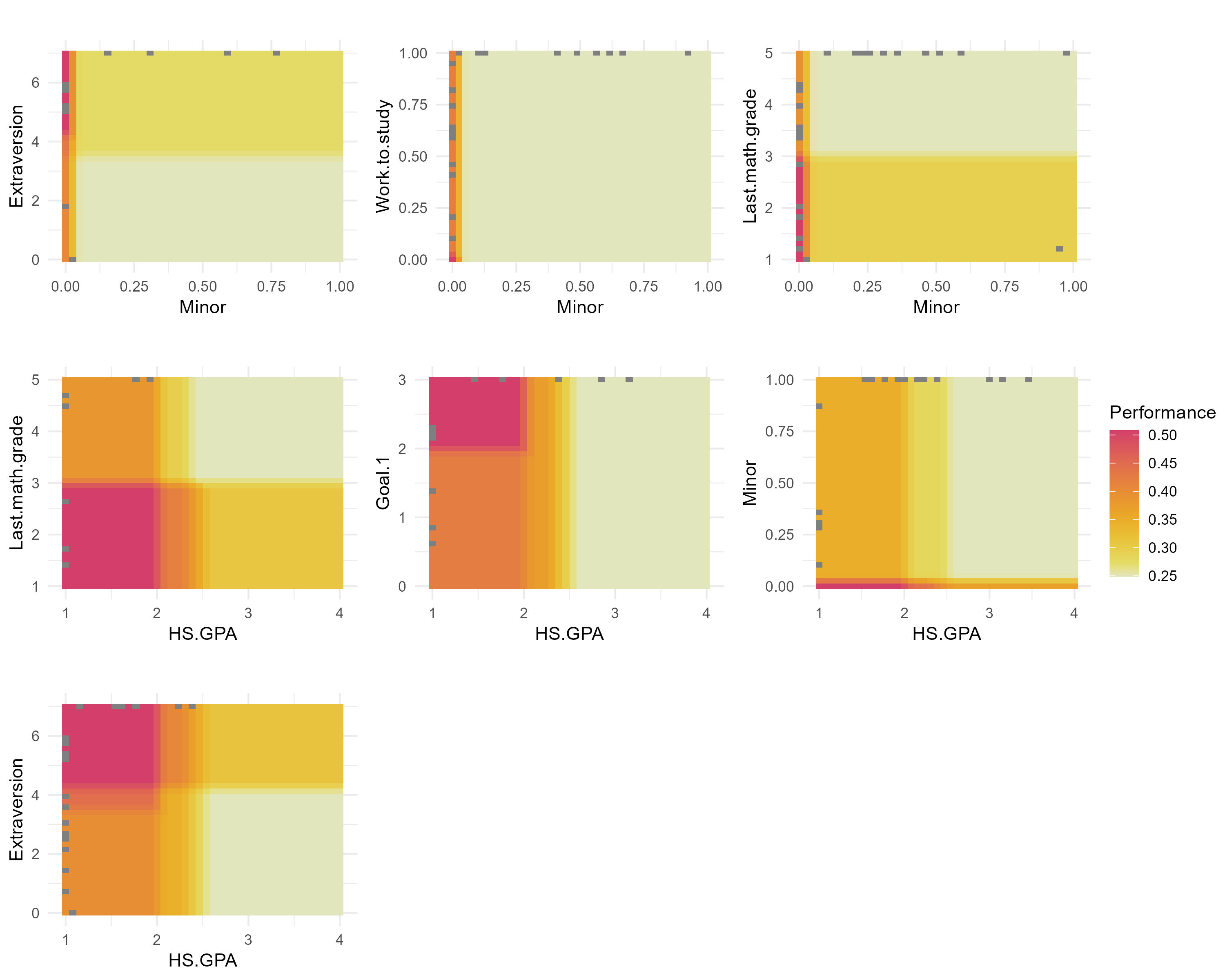}
\end{figure}

\begin{figure}[H]
    \centering
\caption{Bivariate dependencies for all rules selected on practice test performance in the illustrative example for the model trained after using missRanger.}
    \label{fig:bivDepMiss}    
    \includegraphics[width=1\linewidth]{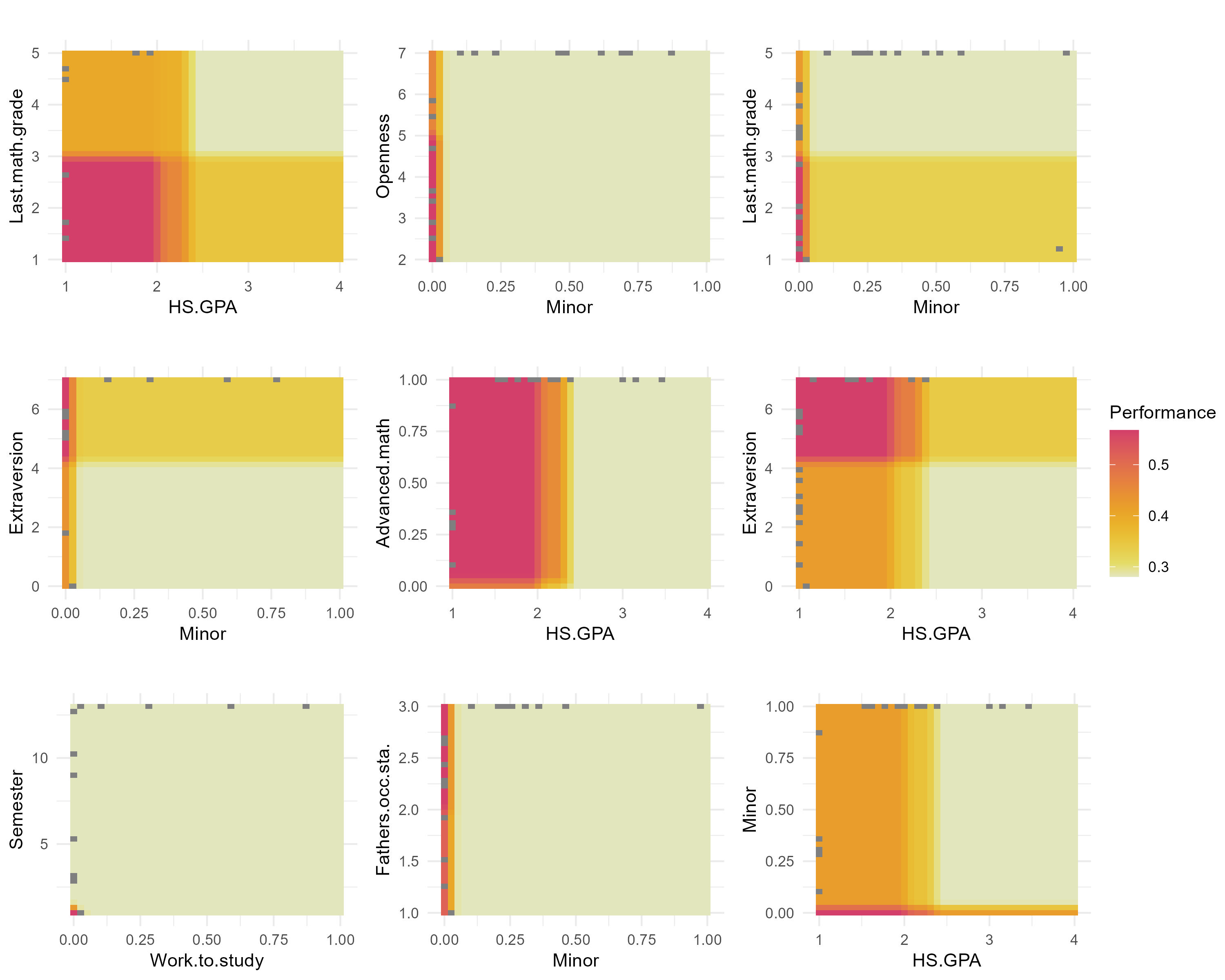}
\end{figure}

\begin{figure}[H]
    \centering
\caption{Bivariate dependencies for all rules selected on practice test performance in the illustrative example for the model trained after using MIXGBoost.}
    \label{fig:bivDepMix}    
    \includegraphics[width=1\linewidth]{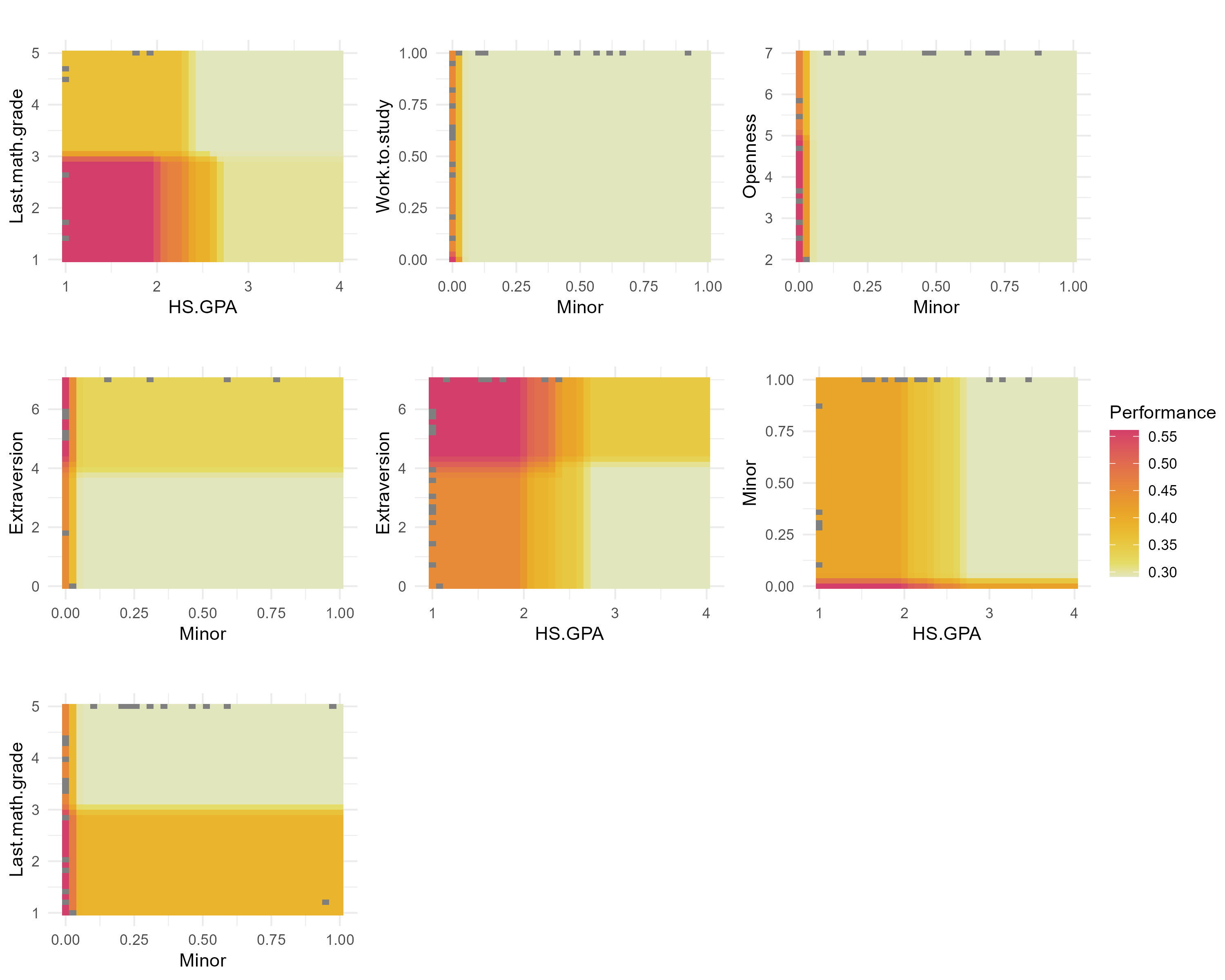}
\end{figure}

\end{document}